\documentclass[preprint,superscriptaddress,showkeys,showpacs,
tightenlines,nofootinbib,byrevtex,amsmath,prd,fleqn]{revtex4}   
\usepackage{graphicx}
\usepackage{bm}
\begin{document}      
\preprint{YITP-09-01}
\preprint{INHA-NTG-01/09}
\title{Photoproduction of $\Theta^+(1540,1/2^+)$ reexamined\\
with new theoretical information}
\author{Seung-il~Nam}
\email[E-mail: ]{sinam@yukawa.kyoto-u.ac.jp}
\affiliation{Yukawa Institute for Theoretical Physics (YITP), Kyoto
University, Kyoto 606-8502, Japan} 
\author{Hyun-Chul~Kim}
\email[E-mail: ]{hchkim@inha.ac.kr}
\affiliation{Department of Physics, Inha University, Incheon 402-751,
  Korea.}  
\date{\today}
\begin{abstract} 
We reinvestigate the photoproduction of the exotic pentaquark baryon 
$\Theta^+(1540,1/2^+)$ from the $\gamma N\to\bar{K}\Theta^+$ reaction
process within the effective Lagrangian approach, taking into account  
new theoretical information on the $KN\Theta$ and $K^*N\Theta$
coupling strengths from the chiral quark-soliton model ($\chi$QSM).
We also consider the crossing-symmetric hadronic form factor,
satisfying the on-shell condition as well. Due to the sizable vector 
and tensor couplings for the vector kaon, $g_{K^*N\Theta}$ and
$f_{K^*N\Theta}$, which are almost the same with the vector coupling 
$g_{KN\Theta}\approx0.8$ for the pseudoscalar kaon, the $K^*$-exchange
contribution plays a critical role in the photon beam asymmetries.  
\end{abstract} 
\maketitle
\section{Introduction}
Since the LEPS collaboration announced the evidence of the
$\Theta^+$~\cite{Nakano:2003qx}, being motivated by
Ref.~\cite{Diakonov:1997mm} in which its decay width was
predicted to be very small with its mass $1540$
MeV~\cite{Praszalowicz:2003ik}, it intrigued a great deal of
experimental and theoretical works on the $\Theta^+$  (see, for
example, reviews~\cite{Hicks:2005gp,Goeke:2004ht} for the experimental
and theoretical status before 2006).  However, the CLAS collaboration 
conducted a series of experiments and reported eventually null results
of finding the $\Theta^+$~\cite{CLAS1,CLAS2,CLAS3,CLAS4} in various
reactions.  Considering the fact that these CLAS experiments were
dedicated ones with high statistics, these null results from the CLAS
experiment are remarkable and indicate that the total cross sections
for photoproductions of the $\Theta^+$ should be tiny.  In fact,
the $95\%$ confidence level (CL) upper limits on the total cross
sections for the $\Theta^+$ at $1540$ MeV lie mostly  in the range of
${(0.3-0.8)}$ nb~\cite{CLAS1,CLAS2,CLAS4}.  In Ref.~\cite{CLAS3} the
upper limit on the $\gamma d\to\Lambda(1520,3/2^-)\Theta^+$ total
cross section turned out to be about 5 nb in the mass range 
from $1520$ MeV to $1560$ MeV with a $95\,\%$ CL. The KEK-PS-E522 
collaboration~\cite{Miwa:2006if} has carried out the experiment
searching for the $\Theta^+$ via the $\pi^- p \to K^- X$ reaction and
found a bump at around $1530$ MeV but with only  
$(2.5\sim2.7)\,\sigma$ statistical significance.  Moreover, the upper
limit of the $\Theta^+$-production cross section in the $\pi^- p \to
K^- \Theta^+$ reaction was extracted to be $3.9\,\mu\mathrm{b}$.  A
later experiment at KEK (KEK-PS-E559), however, has observed no clear
peak structure for the $\Theta^+$ in the $K^+p\to \pi^+ X$
reaction~\cite{Miwa:2007xk}, giving a $95\,\%$ CL upper limit of
$3.5\,\mu$b/sr for the differential cross section averaged over from
$2^\circ$ to $22^\circ$ in the laboratory system.  This negative
situation is summarized in the 2008 Review of Particle 
Physics by Wohl~\cite{PDG2008}: ``{\em The whole story -- the
discoveries themselves, the tidal wave of papers by theorists and
phenomenologists that followed, and the eventual ``undiscovery'' --
is a curious episode in the history of science.}'' 

In the meanwhile, the DIANA collaboration has continued to search for
the $\Theta^+$ in the $K^+n\to K^0 p$ reaction and has reported a
direct formation of a narrow $pK^0$ peak with mass of $(1537\pm 
2)$ MeV and width of $\Gamma=(0.36\pm0.11)$ MeV~\cite{DIANA2}.
Compared to the former measurement by the DIANA collaboration for the
$\Theta^+$, the decay width was more precisely measured in this new 
experiment~\cite{Barmin:2003vv}, the statistics being doubled.  The
SVD experiment has also reported a narrow peak with the mass,
$(1523\pm2_\mathrm{stat.}\pm3_\mathrm{syst.})$ MeV in the inclusive
reaction $pA\to pK_s^0+X$~\cite{SVD,SVD2008}.  Moreover, the LEPS
collaboration has brought news very recently on the evidence of the
$\Theta^+$~\cite{Nakano:2008ee}: The mass of the $\Theta^+$ is found
at $(1525\pm 2+3)$ MeV and the statistical significance of the peak
turns out to be $5.1\,\sigma$.  The peak position is shifted by
$+3\,\mathrm{MeV}$ systematically due to the minimum momentum
spectator approximation (MMSA).  The differential cross section was
estimated to be $(12\pm2)\, \mathrm{nb/sr}$ in the photon energy
ranging from 2.0 GeV to 2.4 GeV in the LEPS angular range. 

Although it seems that the pentaquark baryon $\Theta^+$ rarely exists
according to certain experiments as explained above, it is still
theoretically necessary to understand why the $\Theta^+$ is so elusive
and intractable.  As mentioned previously, one of the reasons can be
found in the fact that the cross sections of the $\Theta^+$
photoproduction as well as of the mesonic production are observed to
be minuscule.  The origin of these tiny cross sections can be 
understood by the smallness of the $KN\Theta$ and $K^*N\Theta$
coupling constants, as mentioned explicitly in
Refs.~\cite{Miwa:2007xk}.  A similar conclusion was also found in
Ref.~\cite{Nakano:2008ee}.  Moreover, the decay width of the
$\Theta^+\to KN$, observed by the DIANA collaboration, indicates that
the $KN\Theta$ coupling constant must be small, as was reviewed in
Ref.~\cite{Diakonov:2007zz}.  From the theoretical side,
Ref.~\cite{Kwee:2005dz} has shown that the tensor coupling constant
for the $K^*N\Theta$ vertex is very small and has predicted the total
cross section for the $\Theta^+$ photoproduction to be around
$0.2\,\mathrm{nb}$ even before the CLAS measurement, in which the
cross section for the $\Theta^+$ photoproduction was estimated to be
below $(0.3\sim0.8)$ nb~\cite{CLAS1,CLAS2,CLAS4}.  Azimov {\it et
  al.}~\cite{Azimov:2006he} has evaluated the smaller value of the
$K^*N\Theta$ tensor coupling constant, employing the vector meson
dominance with SU(3) symmetry.  The vector coupling constant
for the $K^*N\Theta$ vertex vanishes in SU(3) symmetry due to the 
generalized Ademollo-Gatto theorem as shown in
Ref.~\cite{Ledwig:1900ri}, in which the vector and tensor coupling 
constants for the $K^*N\Theta$ vertex turned out to be very small
within the framework of the chiral quark-soliton model ($\chi$QSM)
with SU(3) symmetry breaking effects taken into account: The vector
coupling constant $g_{K^*N\Theta}=0.74\sim0.87$ and tensor coupling
constant $f_{K^*N\Theta}=0.53\sim1.16$, respectively.  In the same
theoretical framework, the $KN\Theta$ coupling constant was determined
to be $g_{KN\Theta}=0.83$, which leads to the corresponding the decay
width of the $\Theta^+$: $\Gamma_{\Theta\to NK}=0.71$
MeV~\cite{Ledwig:2008rw}.  Note that these results in the $\chi$QSM are
all derived without adjusting any parameters.      

In Refs.~\cite{Nam:2005uq,Liu:2003zi,Oh:2003kw,Nam:2004xt}, the
photoproduction of the $\Theta^+$ was investigated, based on 
effective Lagrangian approaches.  However, since the coupling
constants and cut-off masses were unknown both experimentally and
theoretically, it was very difficult to describe the production
mechanism of the $\Theta^+$ without any ambiguity.  Thus, in the
present work, we want to reexamine the photoproduction of the
$\Theta^+$, incorporating the $KN\Theta$ and $K^*N\Theta$ coupling
cosntants and cut-off masses from
Refs.~\cite{Ledwig:1900ri,Ledwig:2008rw}.  The results will be shown
that the magnitudes of the total cross section and differential cross
section are qualitatively compatible with those of the LEPS and CLAS
data. 

We sketch the structure of the present work as follows: In Section II,
we briefly review the coupling constants of the $K$ and $K^*$
exchange, which play critical roles in describing the photoproduction
of the $\Theta^+$.  In Section III, we explain the general formalism
of the effective Lagrangian method.  In Section IV, we present the
numerical results and discuss them.  The final Section is devoted to
summary and conclusion.     
\section{Coupling constants and form factors for the $K$ and $K^*$   
 exchanges from the  $\chi$QSM}  
In this Section, we briefly review the results of the $\chi$QSM
calculations. We start with the following $\Theta^{+}$-to-neutron 
transition matrix elements of the vector current $V^{\mu} =
\bar{\psi}\gamma^{\mu}\frac{1}{2}(\lambda^{4}-i\lambda^{5})\psi$,   
and axial-vector current $A^{\mu} = \bar{\psi} \gamma^{\mu}
\gamma^{5} \frac{1}{2}(\lambda^{4}-i\lambda^{5})\psi$:
\begin{eqnarray}
\langle\Theta(p^{\prime})|V^{\mu}(0)|n(p)\rangle & = & 
\bar{u}_{\Theta}(\bm{p}^{\prime})\left[F_{1}^{n\Theta}(Q^{2})
  \gamma^{\mu} + \frac{F_{2}^{n\Theta}(Q^{2})
    i\sigma^{\mu\nu}q_{\nu}}{M_{\Theta}+M_{n}} +
  \frac{F_{3}^{n\Theta}(Q^{2}) q^{\mu}}{M_{\Theta} +
    M_{n}}\right]u_{n}(\bm{p})\,\,\,,\label{eq:Vector-current}\\  
\langle\Theta(p^{\prime})|A^{\mu}(0)|n(p)\rangle & = &
\bar{u}_{\Theta}(\bm{p}^{\prime}) \left[G_{1}^{n\Theta}(Q^{2})
  \gamma^{\mu} + G_{2}^{n\Theta}(Q^{2}) q^{\mu} +
  G_{3}^{n\Theta}(Q^{2}) P^{\mu}\right]\gamma^{5}\,
u_{n}(\bm{p})\,\,\,,\label{eq:Axial-Vector-current} 
\end{eqnarray}
where $u_{\Theta(n)}$ denotes the spinor of the $\Theta^{+}$ (neutron)
with the corresponding mass $M_{\Theta(n)}$.  The $Q^2$ stands for the
momentum transfer $Q^2=-q^2=-(p'-p)^2$ and $P$ represents the total
momentum $P=p'+p$.  $F_i^{n\Theta}$ and $G_i^{n\Theta}$ designate real
transition form factors, related to the strong coupling
constants for the $K^{*}$ and $K$ with the help of the
vector-meson dominance (VMD)~\cite{Sakurai:1960ju,Feynman:1973xc} and
Goldberger-Treiman relations. 

In the VMD, the vector-transition current can be expressed as the
$K^*$ current by the current field identity (CFI):  
\begin{equation}
V^{\mu}(x) \; = \; \bar{s}(x) \gamma^{\mu}u(x) = 
\frac{m_{K^{*}}^{2}}{f_{K^{*}}}K^{*\mu}(x)\,,\label{eq:CFI}
\end{equation}
where $m_{K^*}$ and $f_{K^*}$ denote, respectively, the mass of the
$K^*$ meson, $m_{K^{*}}=892$ MeV, and decay constant defined as 
\begin{equation}
f_{K^{*}}^{2}=\frac{m_{K^{*}}^{2}}{m_{\rho}^{2}}f_{\rho}^{2},
\end{equation}
where the decay constant $f_\rho$ for the rho meson can be determined as 
\begin{equation}
f_{\rho}^{2}=\frac{4\pi\alpha^{2}m_{\rho}}{3\,\Gamma_{\rho^0\to e^+e^-}}.
\end{equation}
Here, $\alpha$ denotes the electromagnetic fine-structure
constant. The $f_{K^{*}}$ is determined by using the $\rho$-meson
experimental data  with $m_{\rho}=770$ MeV and $\Gamma_{\rho^0\to
e^+e^-}=(7.02\pm0.11)$ keV~\cite{PDG2008}, for which we get the
values $f_{\rho}\approx4.96$ and $f_{K^{*}}\approx5.71$.  Then, using
the CFI, we can express the $K^*N\Theta$ vertex in terms of the
transition form factors in
Eqs.~(\ref{eq:Vector-current}) and (\ref{eq:Axial-Vector-current}): 
\begin{eqnarray}
\langle\Theta(p^{\prime})|\bar{s}\gamma^{\mu}u|n(p)\rangle & = &
\frac{m_{K^{*}}^{2}}{f_{K^{*}}}\,\frac{1}{m_{K^{*}}^{2}-q^{2}}\,
\langle \Theta(p^{\prime})|K^{*\mu}|n(p)\rangle,\\ 
\langle\Theta(p^{\prime})|K^{*\mu}|n(p)\rangle & = &
\bar{u}_{\Theta}(\bm{p}^{\prime})\left[g_{K^{*}n\Theta}
  \gamma^{\mu} + f_{K^{*}n\Theta}
  \frac{i\sigma^{\mu\nu}q_{\nu}}{M_{\Theta}+M_{n}} +
  \frac{s_{K^{*}n\Theta} q^{\mu}}{M_{\Theta} + M_{n}} \right]
u_{n}(\bm{p}),
\label{eq:VDM2}  
\end{eqnarray}
where the $g_{K^{*}n\Theta}$ and $f_{K^{*}n\Theta}$ denote the vector 
and tensor coupling constants for the $K^{*}N\Theta$ vertex,
respectively. By comparing the Lorentz-structures the strong 
coupling constants can be determined as 
\begin{equation} 
g_{K^{*}n\Theta} \;=\;  f_{K^{*}} F_{1}^{\Theta n}(0),\;\;
f_{K^{*}n\Theta} \;= \; f_{K^{*}}\, F_{2}^{\Theta n}(0)\,.
\label{eq:gf}
\end{equation} 

Using the generalized Goldberger-Treiman relation, we can get the
strong coupling constant $g_{Kn\Theta}$ for the $KN\Theta$ vertex as
follows: 
\begin{equation}
g_{Kn\Theta}=\frac{G_{1}^{\Theta n}(0)\,(M_{\Theta}+M_{n})}{2f_{K}},
\label{GT_generalized} 
\end{equation} 
where $f_{K}\approx1.2f_{\pi}$ stands for the kaon decay constant.

The form factors $F_{1}^{n\Theta}(Q^{2})$, $F_{2}^{n\Theta}(Q^{2})$
and $G_{A}^{\Theta n}(Q^{2})$ of
Eqs.~(\ref{eq:Vector-current}) and (\ref{eq:Axial-Vector-current}) 
can be expressed in terms of the matrix elements of the vector and
axial-vector currents with their time and space components decomposed
in the $\Theta^+$ rest frame as follows:
\begin{eqnarray}
G_{E}^{n\Theta}(Q^{2}) & = & \int\frac{d\Omega_{q}}{4\pi} \langle
\Theta(p^{\prime})| V^{0}(0)|n(p)\rangle\,,\label{eq:GE}\\ 
G_{M}^{n\Theta}(Q^{2}) & = & 3M_{n}\int \frac{d\Omega_{q}}{4\pi}
\frac{q^{i} \epsilon^{ik3}}{i\bm{q}^{2}} \langle
\Theta(p^{\prime})| V^{k}(0)|n(p)\rangle\,\,\,,\label{eq:GM}\\
G_{A}^{n\Theta}(Q^{2}) & = & -\frac{3}{2{\bm{q}}^{2}}
\sqrt{\frac{2M_{\Theta}}{E_{\Theta}+M_{\Theta}}}
\int\frac{d\Omega_{q}}{4\pi} \Big[{\bm{q}} \times \Big({\bm{q}} \times
\langle \Theta(p^{\prime})| {\bm{A}}(0) |n(p)\rangle\Big)\Big]_{z}\,,
\label{eq:GA}
\end{eqnarray} 
 where the electromagnetic-like form factors $G_{E}^{n\Theta}$ and
$G_{M}^{n\Theta}$ are written as 
\begin{eqnarray}
G_{E}^{n\Theta}(Q^{2}) & = & \sqrt{\frac{E_{n}+M_{n}}{2M_{n}}}
\left[F_{1}^{n\Theta}(Q^{2}) -
  \frac{F_{2}^{n\Theta}(Q^{2})}{M_{\Theta}+M_{n}}
  \frac{{\bm{q}}^{2}}{E_{n}+M_{n}} + F_{3}^{n\Theta}(Q^{2}) 
\frac{q^{0}}{M_{\Theta}+M_{n}}\right]\,,
\label{eq:GE in Fi}\\
G_{M}^{n\Theta}(Q^{2}) & = & \sqrt{\frac{2M_{n}}{E_{n}+M_{n}}}
\left[F_{1}^{n\Theta}(Q^{2}) + F_{2}^{n\Theta}(Q^{2})\right]\,.
\label{eq:GM in Fi}
\end{eqnarray}
Since the second and third parts in Eq.~(\ref{eq:GE in Fi}) turn out to
be very small, we take the following expressions as the vector and
tensor coupling constants: 
\begin{equation}
g_{K^{*}n\Theta} \;=\;  f_{K^{*}} G_{E}^{\Theta n}(0),\;\;
f_{K^{*}n\Theta} \;= \; f_{K^{*}}\, (G_{E}^{\Theta
  n}(0)\,-\,G_{M}^{\Theta n}(0)). 
\label{eq:gf2}
\end{equation}

The next step is to evaluate the form factors of
Eqs.~(\ref{eq:GE}), (\ref{eq:GM}), and (\ref{eq:GA}) within the
self-consistent 
$\chi$QSM.  The model is featured by the following effective low-energy
partition function with quark fields $\psi$ with the number of colors
$N_{c}$ and the pseudo-Goldstone boson field $U(x)$ in Euclidean
space:  
\begin{equation}
\label{eq:part}
\mathcal{Z}_{\mathrm{\chi QSM}} = \int\mathcal{D} \psi\mathcal{D}
\psi^{\dagger}\mathcal{D}U\exp \left[-\int d^{4}x \psi^{\dagger}iD(U)
  \psi\right] = \int\mathcal{D}U\exp( -{\bf \mathcal{S}}_{\mathrm{eff}}[U]),
\end{equation}
\begin{equation}
\label{eq:echl}
{\bf \mathcal{S}}_{\mathrm{eff}}(U) =-N_{c}\mathrm{Tr}\ln iD(U),
\end{equation}
where
\begin{eqnarray}
D(U) & = & \gamma^{4}(i\rlap{/}{\partial} - \hat{m}-MU^{\gamma_{5}}) = 
-i\partial_{4}+h(U)-\delta m,
\label{eq:Dirac}\\
\delta m & = &
\frac{m_{s}-\bar{m}}{3}\gamma^{4}\bm{1}_{3\times3}+
\frac{\bar{m}-m_{s}}{\sqrt{3}} \gamma^{4} \lambda^{8} = M_{1}
\gamma^{4} \bm{1}_{3\times3}+M_{8}\gamma^{4}\lambda^{8}.
\label{eq:deltam}
\end{eqnarray}
The current quark mass matrix is defined as
$\hat{m}=\mathrm{diag}(\bar{m},\,\bar{m},\, 
m_{\mathrm{s}})=\bar{m}+\delta m$.  The $\bar{m}$ designates the
average of the up and down current quark masses with isospin symmetry
assumed.  The $M$ denotes the constituent quark mass of which the best
value for the numerical results is $M=420$ MeV.  The pseudo-Goldstone
boson field $U^{\gamma_5}$ is defined as 
\begin{equation}
U^{\gamma_5} \;=\; \exp(i\gamma_5 \lambda^a \pi^a) \;=\;
\frac{1+\gamma_5}{2} U + \frac{1-\gamma_5}{2} U^\dagger 
\end{equation}
with $U=\exp(i \lambda^a \pi^a)$.  For the quantization, we consider
here Witten's embedding of SU(2) soliton into SU(3):
\begin{equation}
U_{\mathrm{SU(3)}} \;=\; \left(
  \begin{array}{cc}
U_{\mathrm{SU(2)}} & 0 \\ 0 & 1     
  \end{array} \right)
\end{equation}
with the SU(2) hedgehog chiral field
\begin{equation}
U_{\mathrm{SU(2)}} \;=\; \exp[i\gamma_5 \hat{\bm n}\cdot \bm\tau
P(r)],   
\label{eq:hedgehog} 
\end{equation}
Here, the $P(r)$ denotes the profile function of the chiral soliton 
$U_{\mathrm{SU(2)}}$.  

In order to describe the baryonic properties, we first have to derive
the profile function.  It can be obtained by the following procedure:
First, we take the large $N_c$ limit and solve it in the saddle-point
approximation, which corresponds at the classical level to finding the
profile function $P(r)$ in Eq.~(\ref{eq:hedgehog}). Thus, the $P(r)$
can be obtained by solving numerically the calssical equation of
motion coming from $\delta S_{\mathrm{eff}}/\delta P(r) =0$, which
yields a classical soliton field $U_c$ constructed from a set of
single quark energies $E_n$ and corresponding states $|n\rangle$
related to the eigenvalue equation $h(U)|n\rangle =E_n|n\rangle$.
However, the classical soliton does not have the quantum number of the 
baryon states, so that we need to project it to physical baryon states
by the semiclassical quantization of the rotational and translational
zero modes.  Note that the zero modes can be treated exactly within
the functional integral formalism by introducing collective
coordinates.  Detailed formalisms can be found in
Refs.~\cite{Christov:1995vm,Kim_eleff}. Considering the rigid
rotations and translations of the classical 
soliton $U_c$, we can express the soliton field as
\begin{equation}
U(\bm x, t) = A(t)U_c(\bm x - \bm z(t))A^\dagger (t),
\end{equation}
where $A(t)$ denotes a unitary time-dependent SU(3) collective
orientation matrix and $\bm z(t)$ stands for the time-dependent
displacement of the center of mass of the soliton in coordinate
space.

In the $\chi$QSM, the baryon state consists of $N_{c}$ valence quarks
expressed as 
\begin{equation}
|B(p)\rangle=\lim_{x_{4}\to-\infty}\,\frac{1}{\sqrt{\mathcal{Z}}}\,
e^{ip_{4}x_{4}}\,\int d^{3}\vec{x}\, e^{i\,\vec{p}\cdot\vec{x}}\,
J_{B}^{\dagger}(x)\,|0\rangle\,
\end{equation}
with the barynic current: 
\begin{equation}
J_{B}(x)=\frac{1}{N_{c}!}\,\Gamma_{B}^{\alpha_{1}\cdots \alpha_{N_{c}}}\, 
\varepsilon^{i_{1}\cdots i_{N_{c}}}\,
\psi_{\alpha_{1} i_{1}}(x)\cdots\psi_{\alpha_{N_{c}} i_{N_{c}}}(x)\,,
\end{equation}  
where $\alpha_{1},\cdots, \alpha_{N_c}$ and $i_1, \cdots, i_{N_c}$ denote
the spin-flavor and color indices, respectively.  The
$\Gamma_B^{\alpha_{1} \cdots \alpha_{N_{c}}}$ stands for the projection
operator for the corresponding baryon state.  Thus, the transition
matrix elements in Eqs.~(\ref{eq:GE}), (\ref{eq:GM}), and (\ref{eq:GA}) can be
written as the following correlation functions:
\begin{eqnarray}
 &  & \langle B_{2}(p_{2})|\mathcal{J}^{\mu\chi}(0)|B_{1}(p_{1}\rangle
 = \frac{1}{\mathcal{Z}} \lim_{T\to\infty}
 e^{-ip_{2}^{4}\frac{T}{2}+ip_{1}^{4}\frac{T}{2}} \int d^{3}
 \vec{x}^{\prime} d^{3} \vec{x}
 e^{i\vec{p}_{1}\cdot\vec{x}-i\vec{p}_{2}\cdot\vec{x}^{\,\prime}} 
\nonumber \\
 &  & \times\int\mathcal{D}U\mathcal{D} \psi^{\dagger} \mathcal{D}
 \psi J_{B^{\prime}} \left(\frac{T}{2},\vec{x}^{\prime} \right)
 \mathcal{J}^{\mu\chi}(0) J_{B}^{\dagger} \left(-\frac{T}{2},\vec{x}
 \right) \,\,\exp{\left[-\int
     d^{4}x\,\psi^{\dagger}iD(U)\psi\right]}\,.
\label{eq:corr}
\end{eqnarray}
We can solve Eq.~(\ref{eq:corr}) in the saddle-point approximation
justified in the large $N_c$ limit, taking into account the zero-mode
quantization explained before.  We consider only the rotational
$1/N_c$ corrections and linear $m_{\mathrm{s}}$ corrections. Thus, we
expand the quark propagators in Eq.~(\ref{eq:corr}) with respect to
$\Omega$ and $\delta m$ to the linear order and
$\dot{T}_{z(t)}^{\dagger}T_{z(t)}$ to the zeroth order.

Having carried out a tedious but straightforward
calculation~(see Refs.~\cite{Christov:1995vm,Kim_eleff} for details),
we finally can express the baryonic matrix elements in
Eqs.~(\ref{eq:GE}), (\ref{eq:GM}), and (\ref{eq:GA}) 
as a Fourier transform in terms of the
corresponding quark densities and collective wave-functions of the
baryons:
\begin{equation}
\langle B^{'}(p^{'})| \mathcal{J}_{\mu}^{\chi}(0) |B(p)\rangle =
\int d A \int d^{3}z\,\, e^{i{\bm q}\cdot{\bm z}}\,
\Psi_{B^{\prime}}^{*}(A) \mathcal{F}_{\mu}^\chi({\bm z})\Psi_{B}(A),
\label{eq:model}
\end{equation}
where $\Psi(A)$ denote the collective wavefunctions and
$\mathcal{F}_{\mu}^{\chi}$ represents the quark densities
corresponding to the current operator $\mathcal{J}_\mu^\chi$.  Using
the collective wavefunctions and the quark densities, we immediately 
obtain the transition vector and axial-vector form factors
$G_{E,M,A}^{n\Theta}$ in Eqs.~(\ref{eq:GE}), (\ref{eq:GM}), and (\ref{eq:GA}).  
The corresponding results can be found in
Refs.~\cite{Ledwig:1900ri,Ledwig:2008rw}, which are summarized in
Table~\ref{tab:coup}.   
\begin{table}[b]
\begin{tabular}{ccc|ccc}
\hline
\multicolumn{3}{c|}{$m_{\mathrm{s}}=0$}&
\multicolumn{3}{c}{$m_{\mathrm{s}}=180$ MeV}\\
\hline
$g_{K^*N\Theta}$ & $f_{K^{*}N\Theta}$ & $g_{KN\Theta}$&
$g_{K^*N\Theta}$ & $f_{K^{*}N\Theta}$ & $g_{KN\Theta}$ \\
\hline
$0$ & $2.91$ & $1.41$&
$0.81$ & $0.84$ & $0.83$ \tabularnewline
\hline
\end{tabular}
\caption{The results for the $K^*N\Theta$ and $KN \Theta^+$ coupling 
constants at $Q^2=0$ with and without $m_{\mathrm{s}}$ corrections.
The constituent quark mass $M$ is taken to be $M=420$ MeV.}
\label{tab:coup}
\end{table}
The vector coupling constant $g_{K^* n\Theta}$ vanishes in SU(3)
symmetric case because of the generalized Ademollo-Gatto
theorem.  Note that even the SU(3) symmetry breaking
effects from the Hamiltonian does not contribute to the $g_{K^*
  n\Theta}$. The value of $g_{K^* n\Theta}$ with SU(3) symmetry
breaking comes solely from the wavefunction corrections.  The coupling 
constants for the proton can be obtained easily by considering isospin
factors. 

Note that there is a sign difference in the coupling constants for the
neutron and proton: $g_{K^*n\Theta}=-g_{K^*p\Theta}$ and the same for
the $f_{K^*N\Theta}$~\cite{Ledwig:2008rw}.  However, as shown in
the next section, since the $K^*$-exchange contribution in the
$\Theta^+$-photoproduction provides a $90^\circ$ phase difference from
others, these sign differences for the neutron and proton targets do
not make any difference at all in describing physical observables.   
\section{An effective Lagrangian approach}
We now proceed to calculate the amplitudes for the reaction of the
$\Theta^+$ photoproduction, taking the results of the coupling
constants in Section II as numerical inputs.  
We first define the relevant effective interactions to compute the 
$\Theta^+$ photoproduction in the Born approximation.  Since the
coupling constants derived from the $\chi$QSM are for the $\Theta^+$
with positive parity, we assume here the parity of the $\Theta^+$ to
be positive. 
\begin{eqnarray}
\label{eq:1}
\mathcal{L}_{KN\Theta}&=&
-ig_{KN\Theta}\bar{\Theta}\gamma_{5}KN+ {\rm h.c.},     
\nonumber\\
\mathcal{L}_{K^*N\Theta}&=&-g_{K^*N\Theta}
\bar{\Theta}\gamma_{\mu}K^{*\mu}
N-\frac{f_{K^*N\Theta}}{M_{\Theta}+M_{N}}
\bar{\Theta}\sigma_{\mu\nu}\partial^{\nu}K^{*\mu}N+ {\rm h.c.},
\nonumber\\
\mathcal{L}_{\gamma KK}&=&
ie_K[(\partial^{\mu}K^\dagger)K-(\partial^{\mu}K)K^\dagger]
A_{\mu}+ {\rm h.c.},
\nonumber\\
\mathcal{L}_{\gamma {K} K^{*}}&=& g_{\gamma K K^{*}}
\epsilon_{\mu\nu\sigma\rho}(\partial^{\mu}A^{\nu})
(\partial^{\sigma}K^{\dagger}){K}^{*\rho} 
+ {\rm h.c.},
\nonumber\\
\mathcal{L}_{\gamma NN}
&=&
-e_N\bar{N}\left[\gamma_{\mu}
-\frac{\kappa_N}{2M_N}
\sigma_{\mu\nu}F^{\mu\nu}\right]N+ {\rm h.c.},
\nonumber\\
\mathcal{L}_{\gamma\Theta \Theta}
&=&
-e_\Theta\bar{\Theta}\left[\gamma_{\mu}
-\frac{\kappa_{\Theta}}{2M_{\Theta}}
\sigma_{\mu\nu}F^{\mu\nu}\right]\Theta+ {\rm h.c.},
\end{eqnarray}
where $\Theta$, $N$, $K$, and $K^*_\mu$ denote the fields of the 
$\Theta^+$, the nucleon, the pseudoscalar kaon, and the vector kaon,
respectively.  The $A_\mu$ represents the photon field, whereas the
$F_{\mu\nu}$ the antisymmetric electromagnetic field strength.  The
$g_{\gamma KK^*}$ designates the $\gamma KK^*$ coupling constant that
can be determined by using the experimental data for the $K^*\to
K\gamma$ decay, which yields $0.388\,\mathrm{GeV}^{-1}$ for the
neutral decay and $0.254\,\mathrm{GeV}^{-1}$ for the charged one.  The
$e_K$, $e_N$, and $e_\Theta$ are unit electric charges for the
kaon, nucleon, and $\Theta^+$, 
respectively.  The $\kappa_N$ and $\kappa_\Theta$ stand for the
anomalous magnetic moments of the nucleon and $\Theta^+$,
respectively.  Since the magnetic moment of the $\Theta^+$ is
theoretically known to be rather
small~\cite{Kim:2003ay,Huang:2003bu,Yang:2004jr,Goeke:2004ht}, we take
it to be $\kappa_\Theta\approx -0.8$ as in Ref.~\cite{Nam:2004xt}.  The 
$\sigma_{\mu\nu}$ is a usual antisymmetric spin operator
$\sigma_{\mu\nu}=i[\gamma_\mu,\, \gamma_\nu]/2$.  The $M_N$ and
$M_\Theta$ correspond to the nucleon and $\Theta^+$ masses, and are
taken to be $939$ MeV and $1540$ MeV, respectively.   

Having performed straightforward manipulations, we arrive at the
following invariant amplitudes for  the $s$-  and $u$-channel
contributions, and the $K$- and $K^*$-exchange ones in the
$t$-channel as follows:   
\begin{eqnarray}
\label{eq:2}
&&i\mathcal{M}_s=-g_{KN\Theta}\bar{u}(p_2)\left[e_N\gamma_5
\frac{F_c(\rlap{/}{p}_1+M_N)+F_s\rlap{/}{k}_1}
{s-M^2_N}\rlap{/}{\epsilon}-\frac{e_Q\kappa_N}{2M_N}\gamma_5
\frac{F_s(\rlap{/}{k}_1+\rlap{/}{p}_1+M_N)}
{s-M^2_N}\rlap{/}{\epsilon}\rlap{/}{k}_1\right]u(p_1),
\nonumber\\
&&i\mathcal{M}_u=
-g_{KN\Theta}\bar{u}(p_2)\left[e_\Theta\rlap{/}{\epsilon}
\frac{F_c(\rlap{/}{p}_2+M_{\Theta})-F_u\rlap{/}{k}_1}
{u-M^2_{\Theta}}
\gamma_5-\frac{e_Q\kappa_{\Theta}}{2M_{\Theta}}\rlap{/}{\epsilon}\rlap{/}{k}_1
\frac{F_u(\rlap{/}{p}_2-\rlap{/}{k}_1+M_{\Theta})}
{u-M^2_{\Theta}}\gamma_5\right]u(p_1),
\nonumber\\
&&i\mathcal{M}^K_t=
2e_Kg_{KN\Theta}\bar{u}(p_2)\gamma_{5}
\frac{(k_2\cdot\epsilon)}{t-M^2_K}u(p_1)F_c,
\nonumber\\
&&\mathcal{M}^{K^*}_t=ig_{\gamma K
K^*}\epsilon_{\mu\nu\sigma\rho}\bar{u}(p_2)\Bigg[
g_{K^*N\Theta}(t)\frac{k^{\mu}_1\epsilon^{\nu}k^{\sigma}_2
\gamma^{\rho}}{t-M^2_{K^*}}
\nonumber\\
&&\hspace{5cm}-\frac{f_{K^*N\Theta}(t)}
{2(M_N+M_\Theta)}\frac{k^{\mu}_1\epsilon^{\nu}k^{\sigma}_2
\gamma^{\rho}\rlap{/}{q}_t-\rlap{/}{q}_tk^{\mu}_1\epsilon^{\nu}k^{\sigma}_2
\gamma^{\rho}}{t-M^2_{K^*}}\Bigg]u(p_1)F^{1/2}_v,
\end{eqnarray}
where $\bar{u}$ and $u$ are the Dirac spinors of $\Theta^+$ and
the nucleon.  The four momenta $p_1$, $p_2$, $k_1$, and $k_2$ stand
for those of the nucleon, the $\Theta^{+}$, the photon, and the kaon,
respectively.  The $s$, $u$ and $t$ represent the Mandelstam
variables. Note that in the case of the process $\gamma p \rightarrow
\bar{K}^0\Theta^+$, there is no contribution from the $K$-exchange
contribution.  Moreover, the $K^*$-exchange contribution gives a
  $90^\circ$ phase difference from others as mentioned previously. 

The form factors as functions of the Mandelstem variables are defined
as follows~\cite{Haberzettl:1998eq,Davidson:2001rk}:   
\begin{equation}
\label{eq:FF}
F_{s,u,t}=\frac{\Lambda^4}{\Lambda^4+[(s,u,t)-M^2_{s,u,t}]^2},
\hspace{0.5cm}
F_v=\frac{\Lambda^4}{\Lambda^4+(t-M^2_{K^*})^2},
\end{equation}
where the four-dimensional cutoff mass $\Lambda$ is chosen to be $650$
MeV which is compatible with those used in the
$\Lambda(1520)$~\cite{Nam:2005uq} and
$\Lambda(1116)$~\cite{Ozaki:2007ka} photoproductions. The common
overall form factor $F_c$ is written as follows: 
\begin{equation}
\label{eq:FFc}
F_c(s,t,u)=1-(1-F_s)(1-F_u)(1-F_t),
\end{equation}
which satisfies the on-shell condition and crossing symmetry. We note
that this form-factor scheme preserves the Ward-Takahashi identity
explicitly.  

As for the coupling constants $g_{K^*N\Theta}$, $f_{K^*N\Theta}$, and
$g_{KN\Theta}$, we take the forms of Eq.~(\ref{eq:gf2}) with the
results of the $\chi$QSM.  The corresponding form factors from
the $\chi$QSM can be parameterized as follows~\cite{Ledwig:1900ri}: 
\begin{equation}
\label{eq:VTFF}
G^{n\Theta}_E(t)=G^0_E\left[\frac{\alpha_E\Lambda^2_E}
{\alpha_E\Lambda^2_E-t}\right]^\alpha+b,\hspace{0.5cm}
G^{n\Theta}_M(t)=G^0_M\left[\frac{\alpha_M\Lambda^2_M}
{\alpha_M\Lambda^2_M-t}\right]^\alpha, 
\end{equation}
where fitting parameters $\alpha_{E,M}$, $\Lambda_{E,M}$, and $b$ are
listed in Table~\ref{table1}.  
\begin{table}[b]
\begin{tabular}{cccc|ccc}\hline 
\multicolumn{4}{c|}{$G^{n\Theta}_E(t)$}&
\multicolumn{3}{c}{$G^{n\Theta}_M(t)$}\\
\hline
$G^0_E$&$\alpha_E$&$\Lambda_E$&$b$&
$G^0_M$&$\alpha_M$&$\Lambda_M$\\
\hline
$0.182$&$9.01$&$0.402$&$-0.04$&$0.286$&$0.851$&$0.559$\\ \hline
\end{tabular}
\caption{Relevant parameters for the electric and magnetic
  vector-transition form factors derived in Eq.~(\ref{eq:VTFF}).}  
\label{table1}
\end{table}
These parameters are determined by using the results with the strange 
quark mass $m_s=180$ MeV and the constituent quark mass $M=420$ MeV.
\section{Numerical results and discussion}
\begin{figure}[t]
\begin{tabular}{cc}
\includegraphics[width=7.5cm]{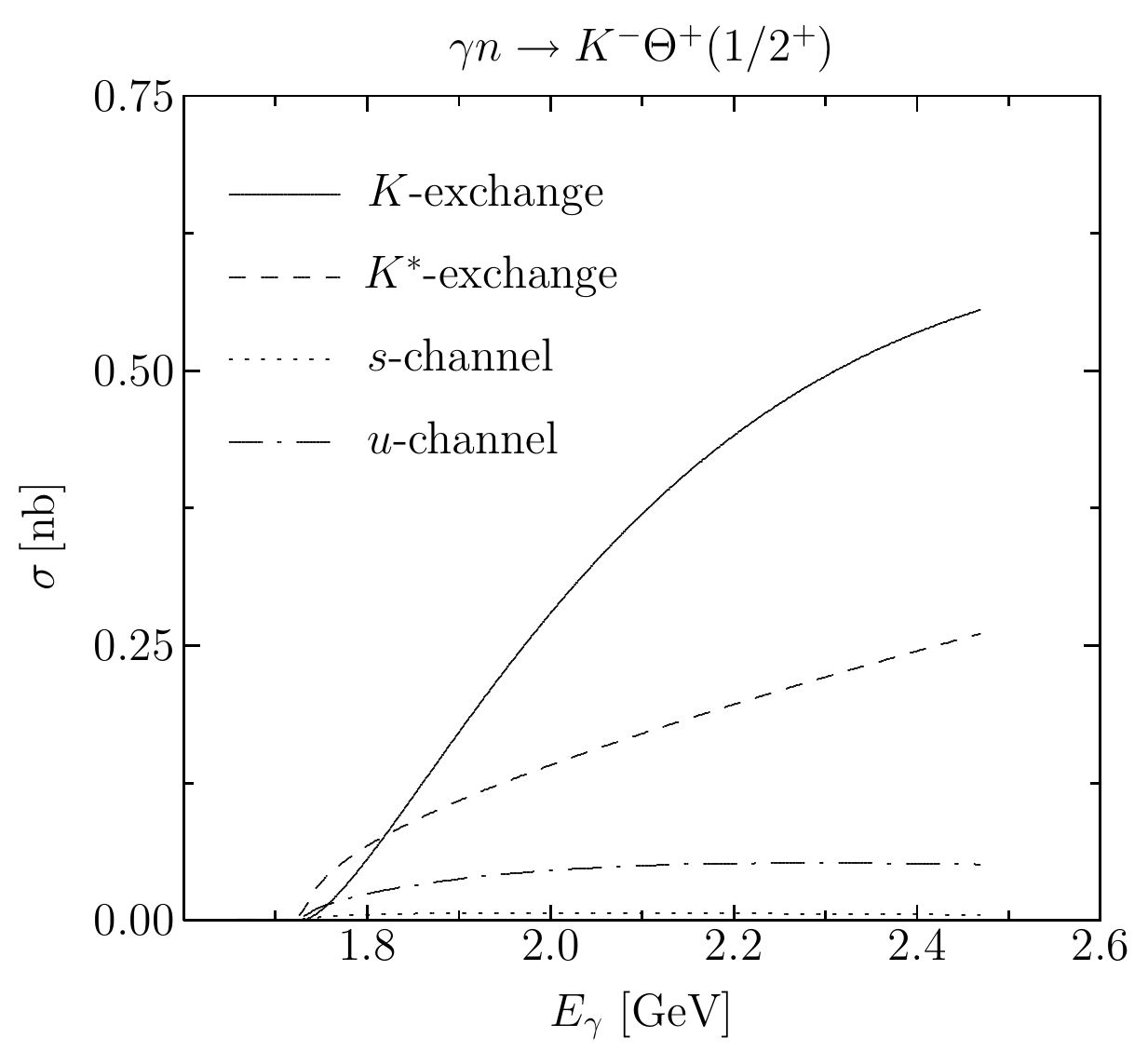}
\includegraphics[width=7.5cm]{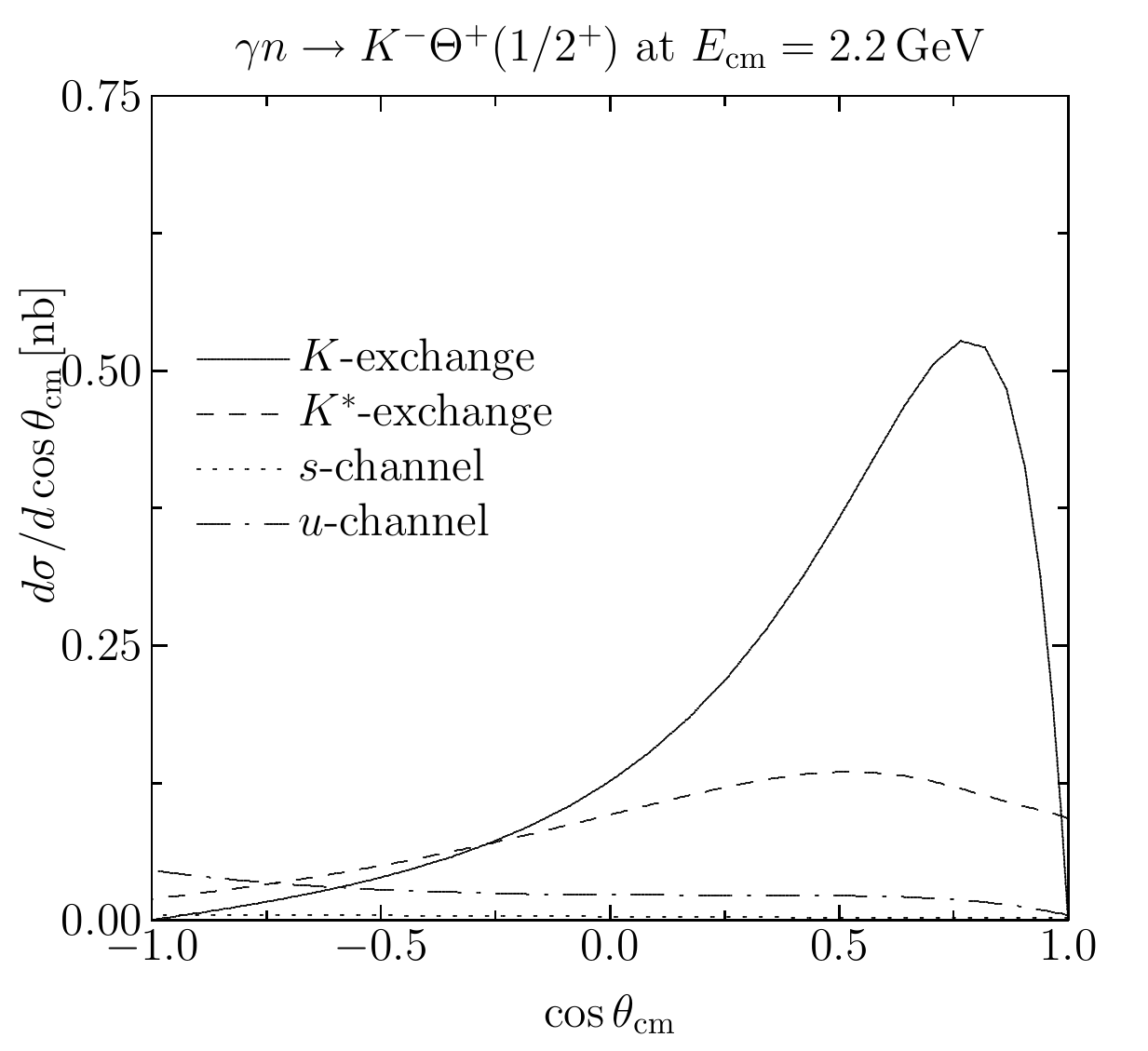}
\end{tabular}
\caption{Each contribution to the total and differential
cross sections for the $\gamma n \to K^- \Theta^+$  reaction. The
total cross section is drawn in the left panel, while the differential
cross section for the photon energy $E_\gamma = 2.2$ GeV is in the
right panel.  The solid curve stands for the  
$K$-exchange contribution, the dashed one for the $K^*$-exchange, the
dotted curve for the $s$-channel, and the dash-dotted one for the
$u$-channel contributions. }
\label{fig:1}
\end{figure}

We are now in a position to discuss the results of the present work.
As mentioned previously, we assume that the $\Theta^+$ baryon has the
spin-parity quantum number, $1/2^+$.  In the left panel of
Fig.~\ref{fig:1}, each contribution to the total 
cross section of the $\gamma n\to K^- \Theta^+$ reaction is drawn as a
function of the photon energy $E_\gamma$.  As shown there in
Fig.~\ref{fig:1}, the $K$-exchange contribution is the most dominant
one, whereas the second dominant one comes from the 
$K^*$-exchange one.  The $K^*$-exchange contribution is, however, the
most dominant one near the threshold up to around $1.8$ GeV, because
of which the total cross section is raised drastically near the
threshold.  The $u$-channel contribution is rather small and the
$s$-channel is almost negligible due to the present form-factor
scheme that suppresses these channels.  The $K$-exchange contribution
shows rather strong dependence on the photon energy $E_\gamma$. As the
$E_\gamma$ increases, it starts to get enhanced. 

The right panel of Fig.~\ref{fig:1} depicts each contribution to the
differential cross section of the $\gamma n\to K^- \Theta^+$ reaction.
The tendency of each contribution is in general similar to the case of
the total cross section, {\it i.e.} the $K$-exchange contribution turns out
to be the dominant one, as it should be.  Moreover, it has a large
bump structure in the forward direction. Beginning from the backward
direction, it is getting increased slowly.  In the forward direction,
it starts to increase drastically till around
$\cos\theta_{\mathrm{cm}}\approx 0.75$ and falls down sharply, which
makes the $K$-exchange contribution have the bump structure. The
$K^*$-exchange contribution is also larger in the forward direction than in the
backward direction.  The $u$-channel contributes mainly to the
backward direction as expected. Summing up all contributions, we can
easily see that the differential cross section is much more enhanced
in the forward direction.
\subsection{Effects of the $K^*$ exchange}
\begin{figure}[t]
\begin{tabular}{cc}
\includegraphics[width=7.5cm]{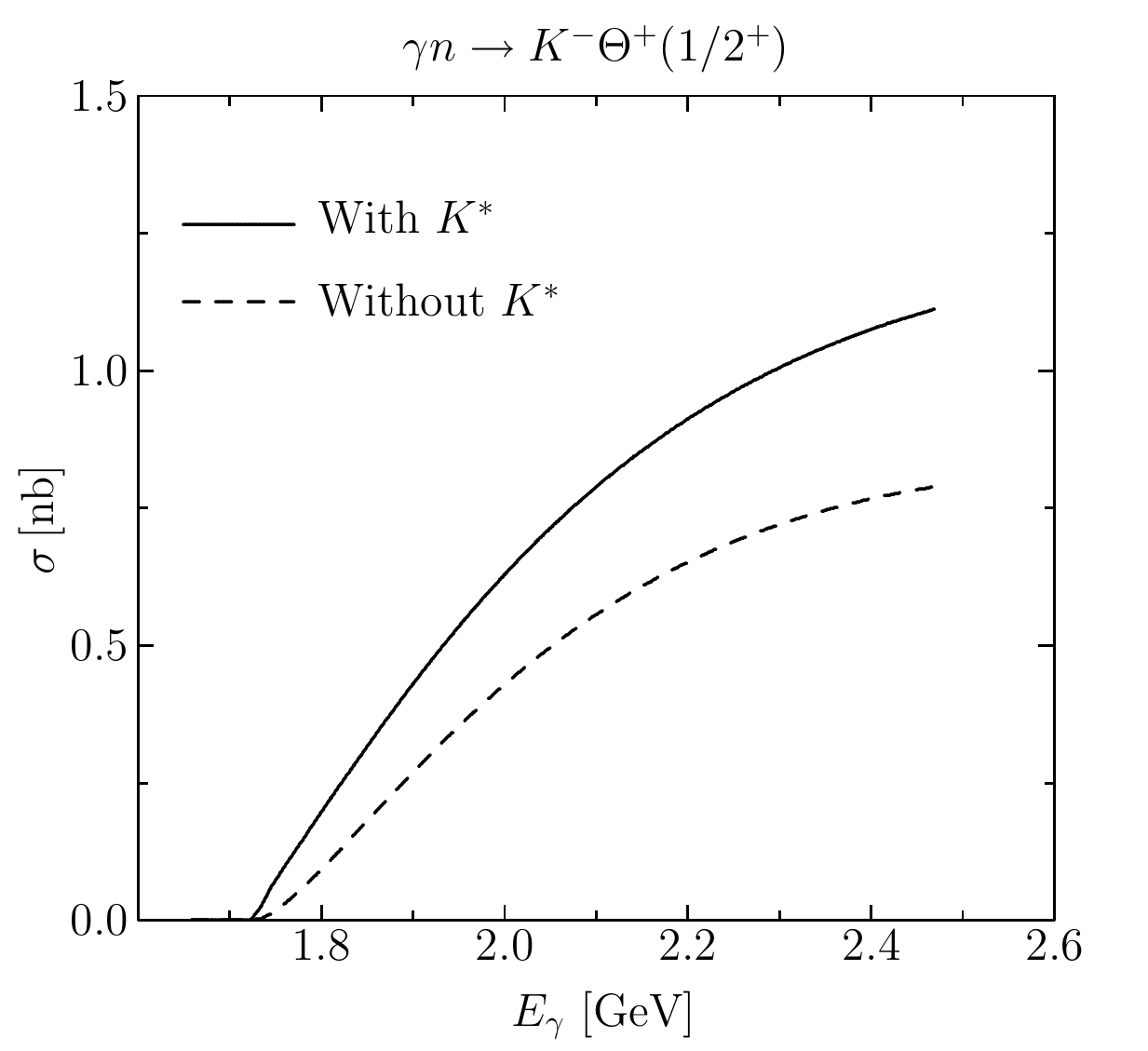}
\includegraphics[width=7.5cm]{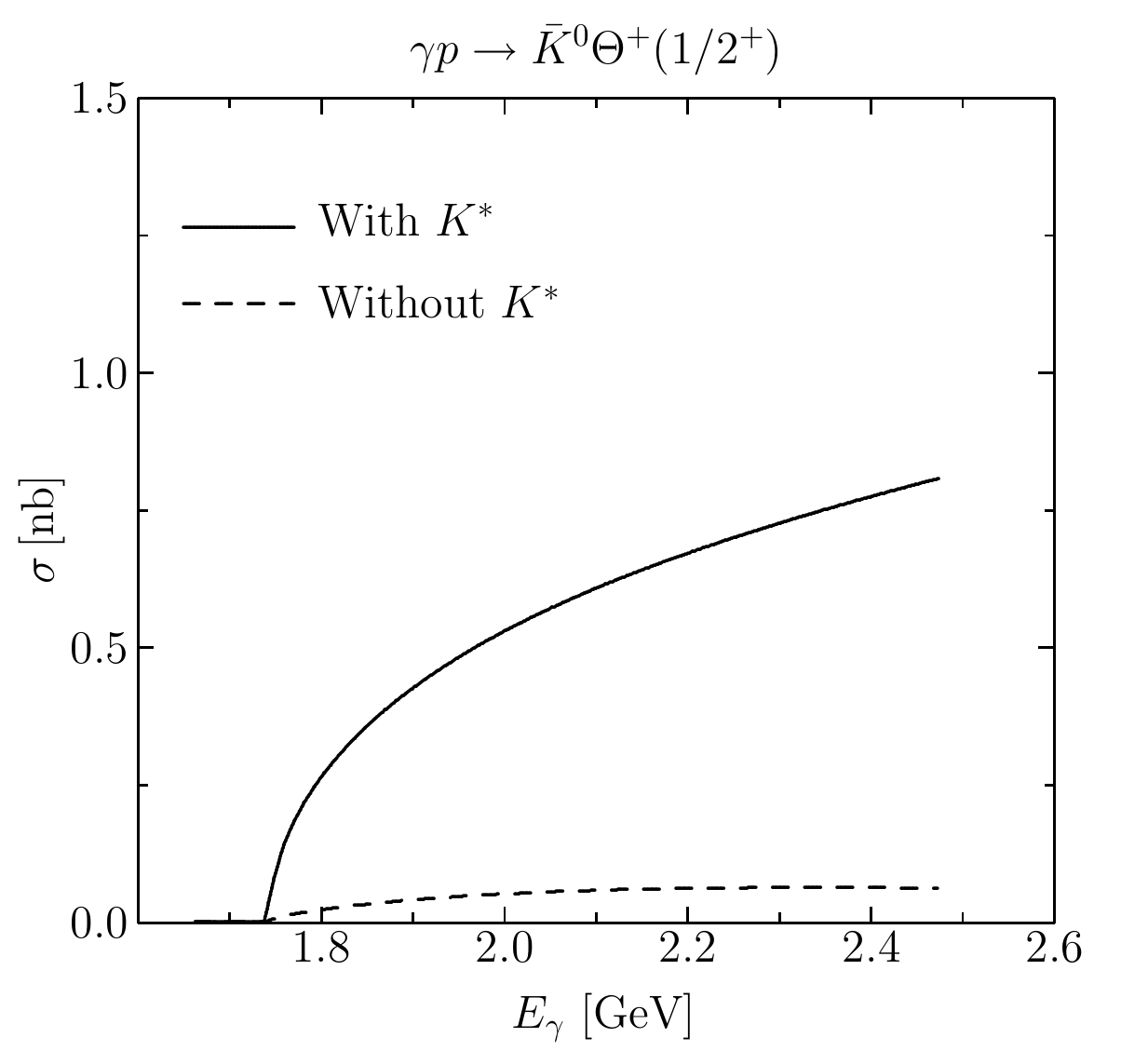}
\end{tabular}
\begin{tabular}{cc}
\includegraphics[width=7.5cm]{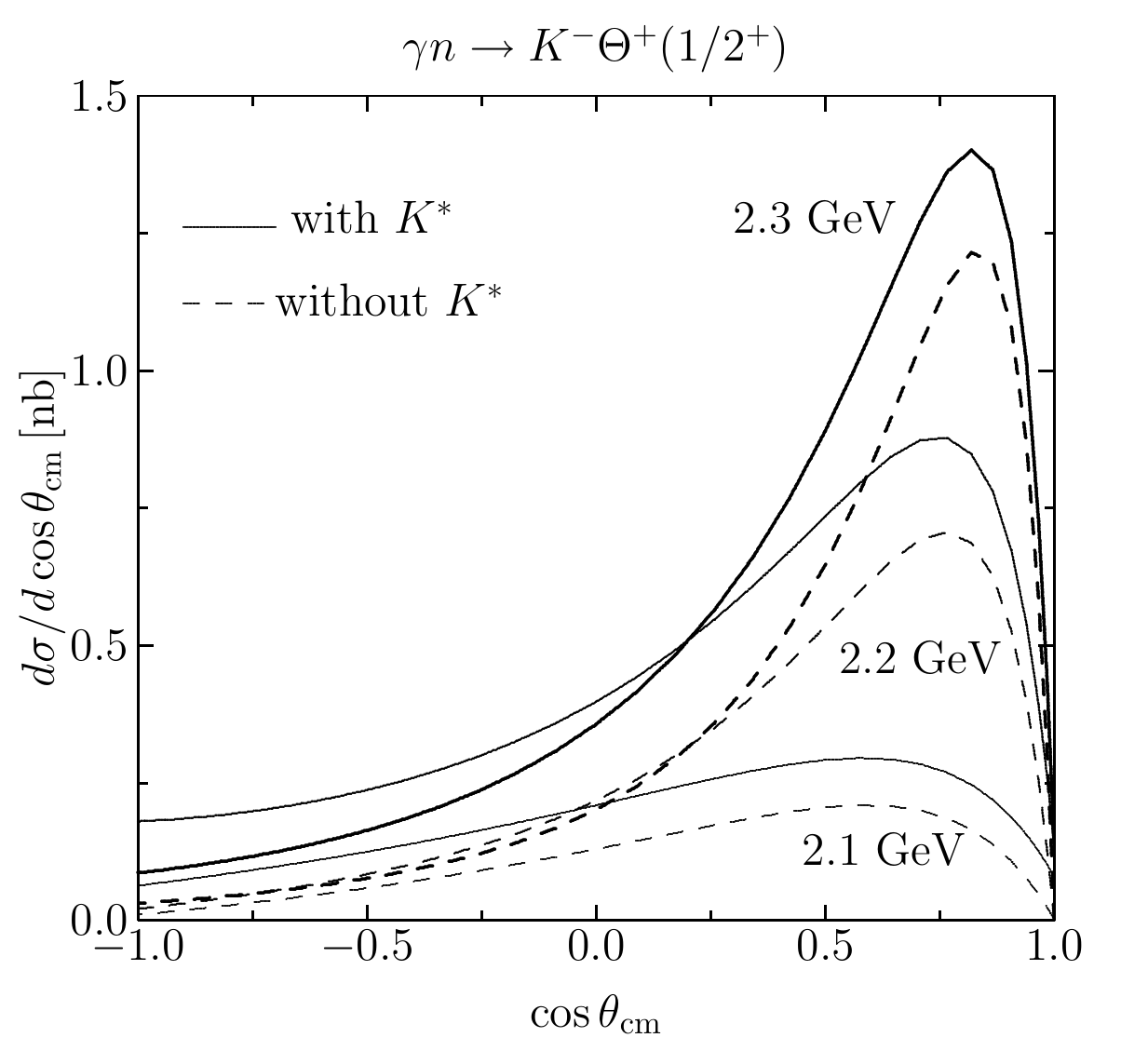}
\includegraphics[width=7.5cm]{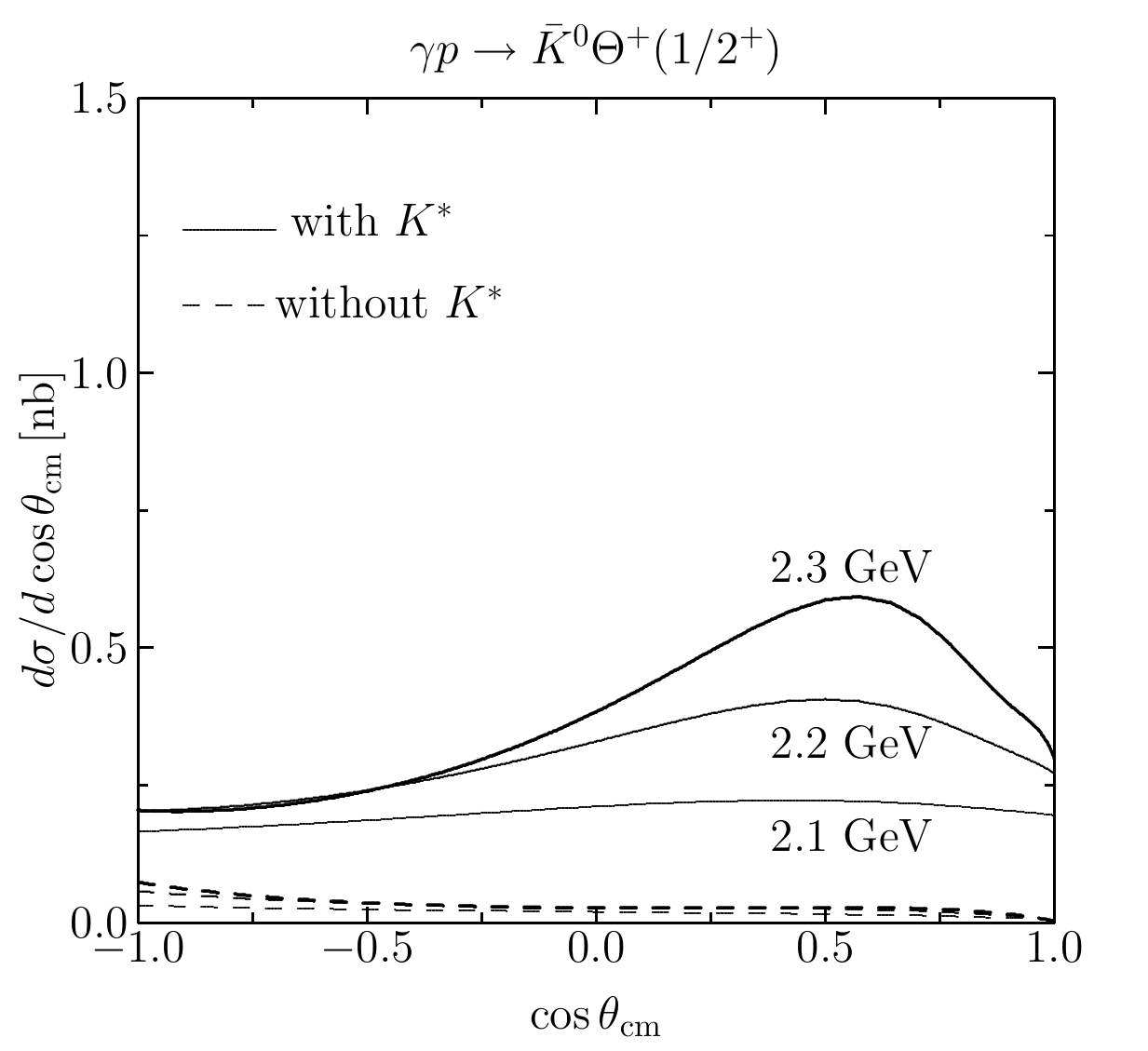}
\end{tabular}
\caption{Effects of the $K^*$-exchange on the total (upper panels) and
differential (lower panels) cross sections. The left panels
represent those for the $\gamma n \to K^- \Theta^+$  reaction, while
the right panels those for the $\gamma p \to \bar{K}^0 \Theta^+$.  The
solid curves indicate those with all contributions, whereas the dashed
one those without the $K^*$-exchange. The differential cross sections
are drawn for three different photon energies $E_\gamma$, $2.1$ GeV,
$2.2$ GeV, and $2.3$ GeV.}  
\label{fig:2}
\end{figure}

Since the present work is mainly interested in the effects of
the $K^*$-exchange contribution, we now examine the features of the
$K^*$-exchange contribution to various observable in detail.  In the
two-upper panels of Fig.~\ref{fig:2}, the total cross sections of
the $\gamma n \to K^- \Theta^+$ and $\gamma p \to \bar{K}^0 \Theta^+$
reactions are drawn in the left and right panels, respectively.  The
$K^*$-exchange contribution turns out to be almost $30\,\%$ to the
total cross section for the neutron target.  On the other hand, it is
almost everything for the proton target.  It can be easily understood
from the fact that for the $\gamma p \to \bar{K}^0 \Theta^+$ reaction
there is no $K$-exchange contribution that is dominant in the neutron
channel. Comparing the total cross sections for the neutron target
with the proton one, we find that that for the neutron one is about
$30\,\%$ larger than that for the proton one, although they are
qualitatively in a similar order $\lesssim1$ nb.

In the two lower panels of Fig.~\ref{fig:2}, we show the
differential cross sections for the $\gamma n \to K^- \Theta^+$ and
$\gamma p \to \bar{K}^0 \Theta^+$ reactions with and without the
$K^*$-exchange contribution for three different photon energies $2.1$
GeV, $2.2$ GeV, and $2.3$ GeV, respectively, in the left and right
panels.  According to the $K$- and $K^*$-exchange contributions,
one can observe the bump structures in the region 
$\lesssim60^\circ$ for both the neutron and proton target cases.  As
in the case of the total cross sections, while the $K^*$-exchange
contribution makes the differential cross section about $10\,\%$
enhanced for the neutron target, its effects are remarkably large for
the proton target. As the photon energy increases, the differential
cross sections also increase consistently, as expected. 

\begin{figure}[t]
\begin{tabular}{cc}
\includegraphics[width=7.5cm]{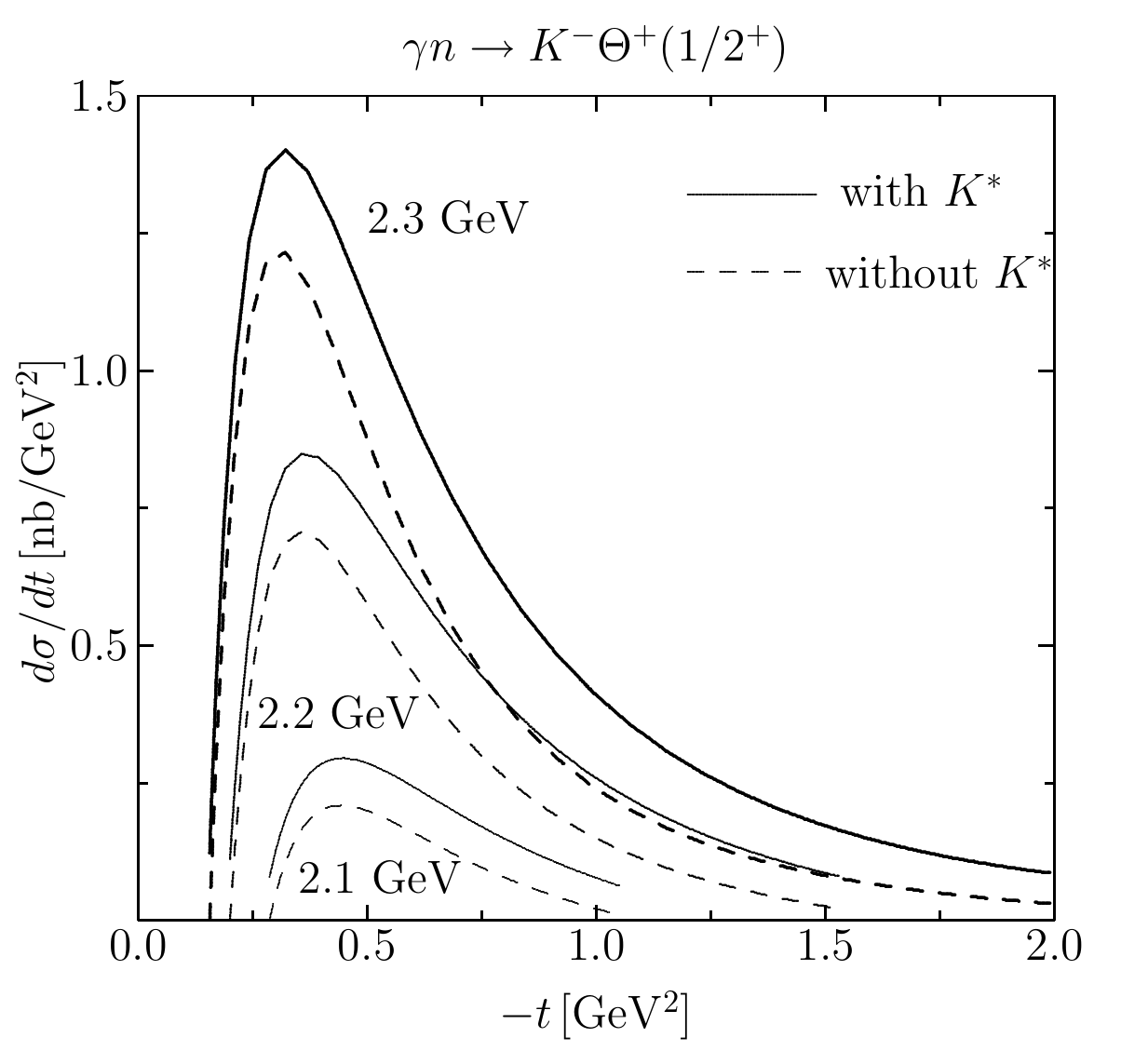}
\includegraphics[width=7.5cm]{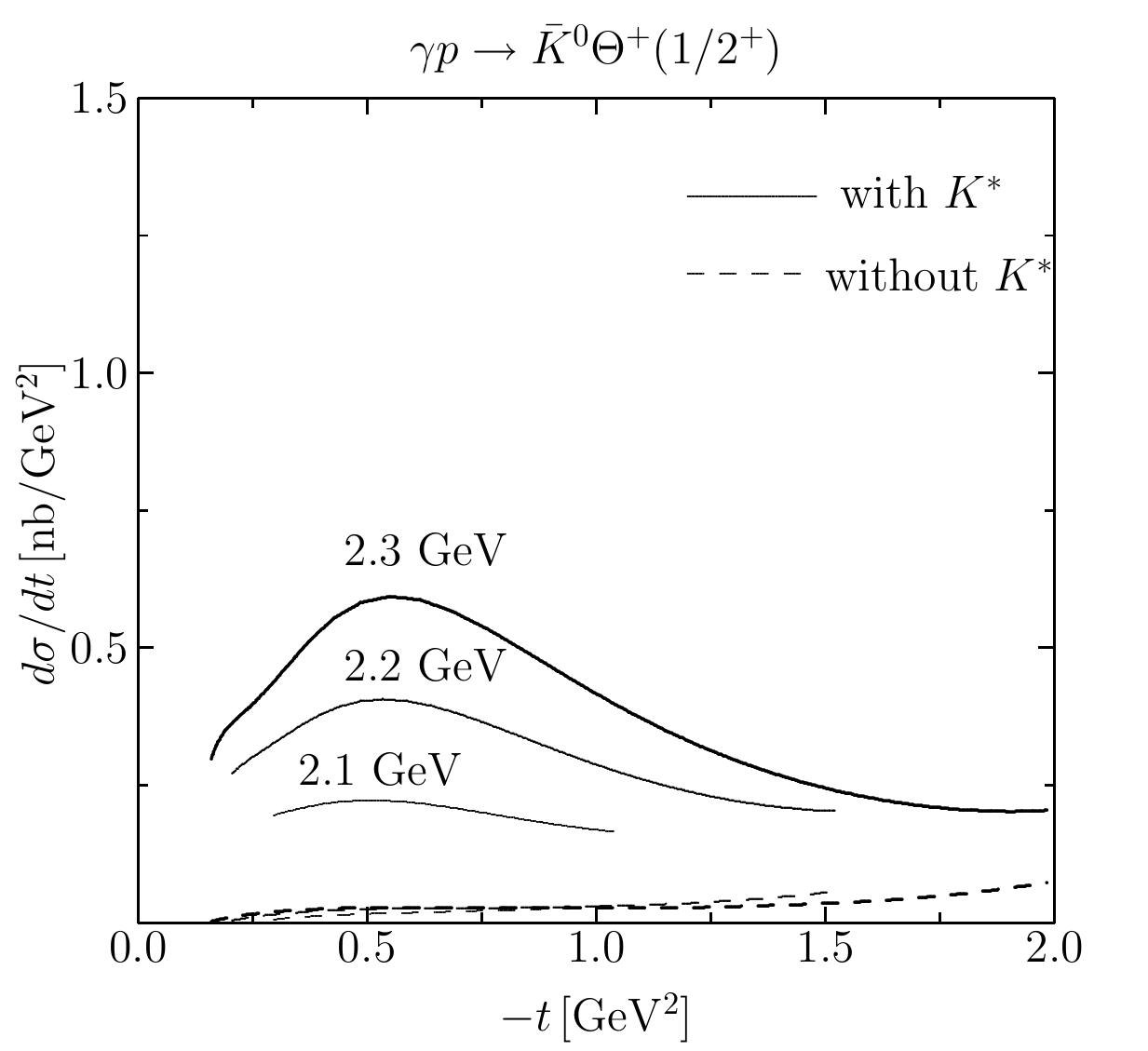}
\end{tabular}
\begin{tabular}{cc}
\includegraphics[width=7.5cm]{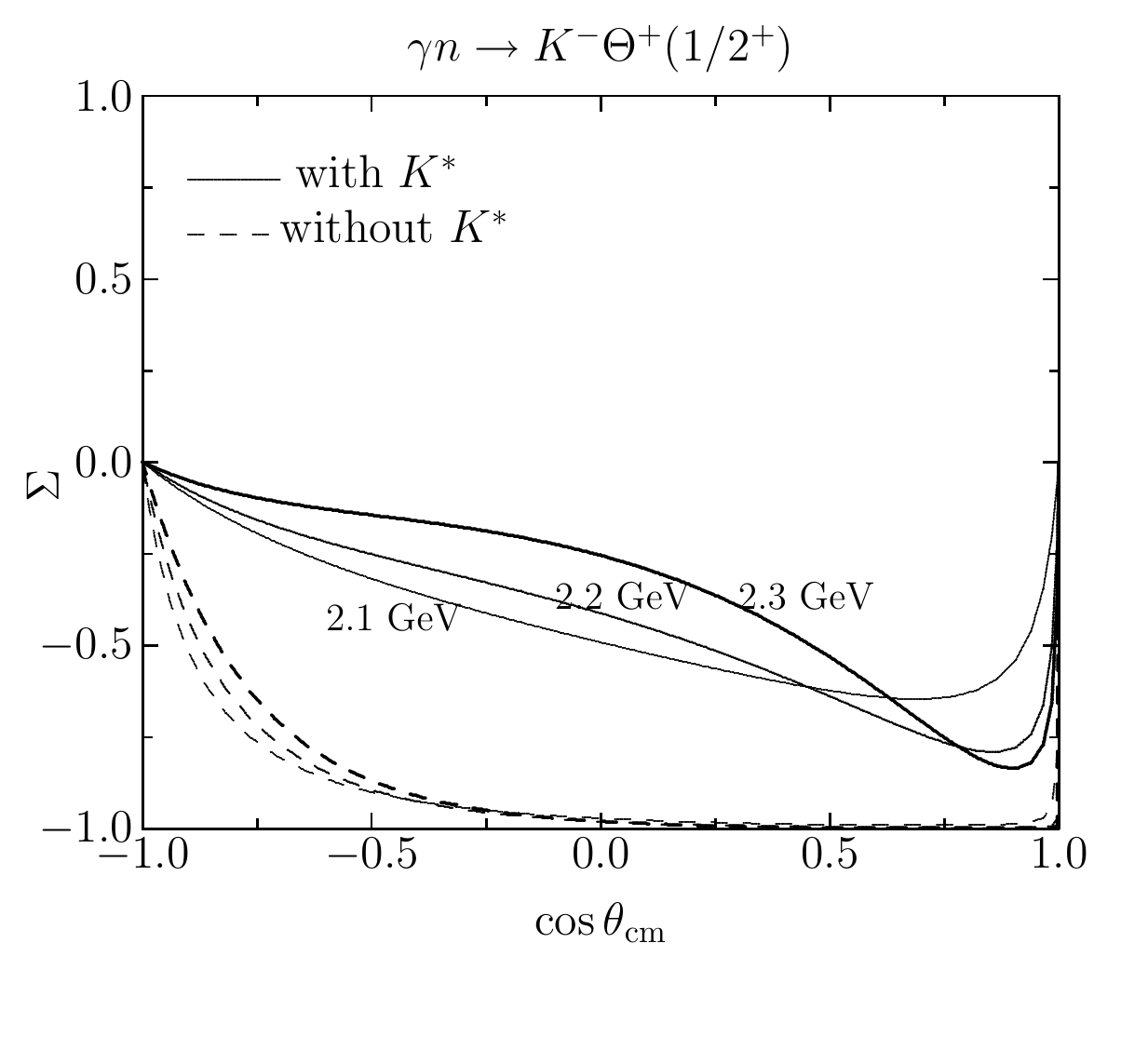}
\includegraphics[width=7.5cm]{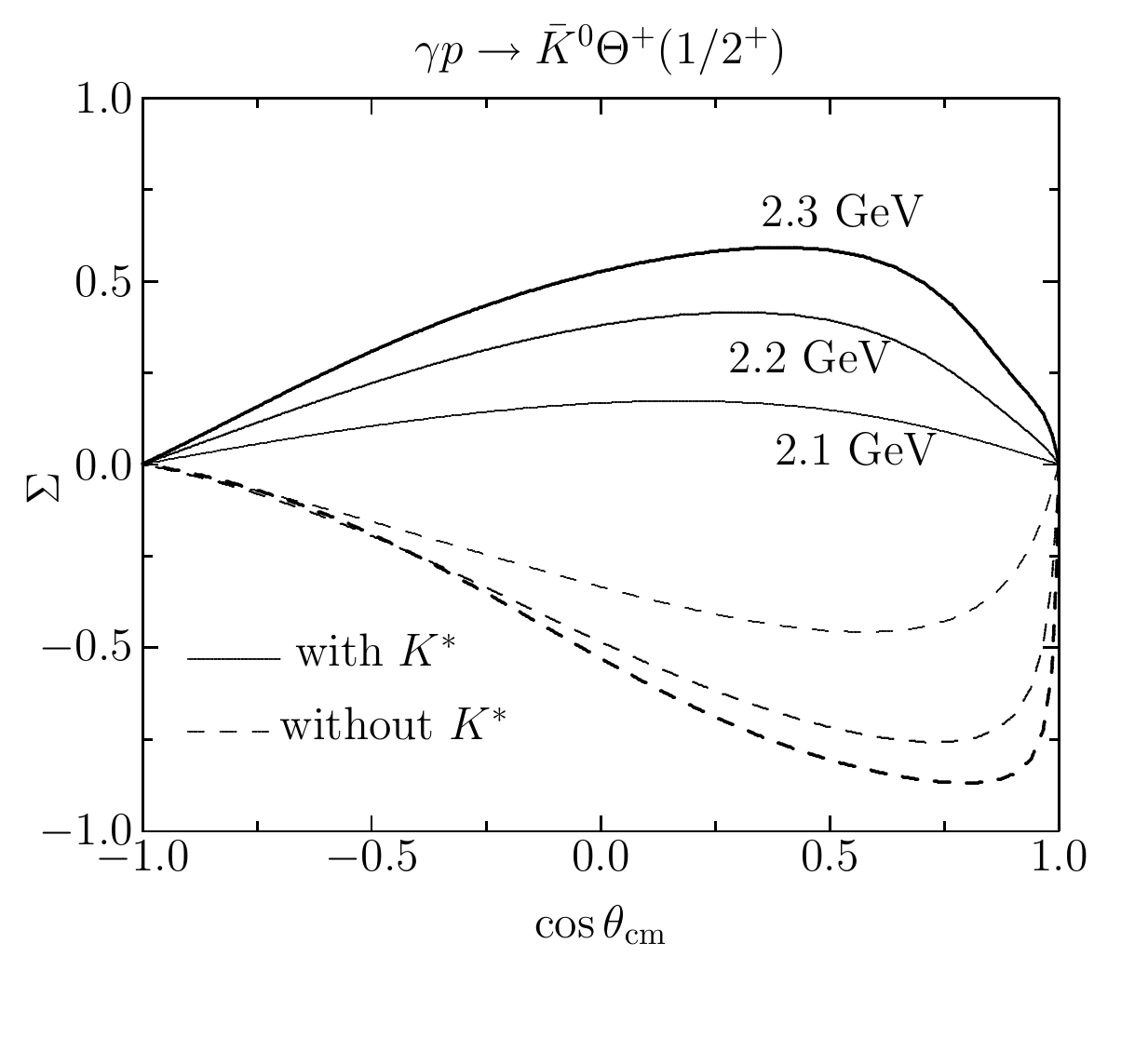}
\end{tabular}
\caption{Effects of the $K^*$-exchange on the $t$-dependences
  ($d\sigma/dt$, upper panels) and the photon-beam asymmetries
  ($\Sigma$, lower panels). The left panels represent those for the  
$\gamma n \to K^- \Theta^+$  reaction, while the right panels those 
for the $\gamma p \to \bar{K}^0 \Theta^+$.  The solid curve indicates
the case with all contributions, whereas the dashed one that without
the $K^*$-exchange.  The curves are drawn for three different photon
energies $E_\gamma$, $2.1$ GeV, $2.2$ GeV, and $2.3$ GeV.} 
\label{fig:3}
\end{figure}

In the two upper panels of Fig.~\ref{fig:3}, we represent the
differential cross section as a function of the momentum transfer $t$.
The general tendency is very similar to that of the differential
cross sections shown in Fig.~\ref{fig:2}.  It is worth
mentioning that the best way to examine the effects of the
$K^*$-exchange contribution is to investigate the photon beam 
asymmetry, since the $K^*$ meson is a vector meson which
manifests magnetic meson-baryon coupling behavior in the present
photoprodcution process.  The photon beam asymmetry is defined as  
\begin{equation}
\Sigma=\left[\frac{d\sigma}{d\Omega}_{\perp}
-\frac{d\sigma}{d\Omega}_{\parallel}\right]\times\left[\frac{d\sigma}
{d\Omega}_{\perp}+\frac{d\sigma}{d\Omega}_{\parallel}\right]^{-1},
\label{eq:BA}
\end{equation}
where the subscript $\perp$ ($\parallel$) denotes that the polarization
vector of the incident photon is perpendicular (parallel) to the
reaction plane.  In the two-lower panels of Fig.~\ref{fig:3} draws the
photon beam asymmetries for the $\gamma n \to K^- \Theta^+$ and
$\gamma p \to \bar{K}^0 \Theta^+$ reactions in the left and right
panels, respectively.  When we switch off the $K^*$-exchange
contribution, the photon beam asymmetry for the neutron target, 
starting from the backward direction, is brought down drastically and
reaches down to almost $\Sigma=-1$ at around
$\theta_{\mathrm{cm}}=90^\circ$, due to the electric meson-baryon
coupling of the dominant $K$-exchange contribution.  However, when we 
turn on the $K^*$-exchange one, the photon beam asymmetry decreases
mildly from the backward direction to the forward direction, and then
it increases sharply to $\Sigma=0$. On the whole, the photon beam
asymmetry is negative for the neutron target.      
  
When it comes to the proton target, the $K^*$-exchange contribution
shows profound effects on the photon beam asymmetry. While the
photon beam asymmetry becomes negative without the $K^*$-exchange
contribution, it turns into being positive with it in all the regions.
With the $K^*$-exchange contribution switched on, the photon beam
asymmetry starts to increase from the bakcward direction to the
forward direction, and it goes down from around
$\cos\theta_{\mathrm{cm}}=0.5$. 
\subsection{Effects of explicit SU(3) symmetry breaking}
\begin{figure}[t]
\begin{tabular}{cc}
\includegraphics[width=7.5cm]{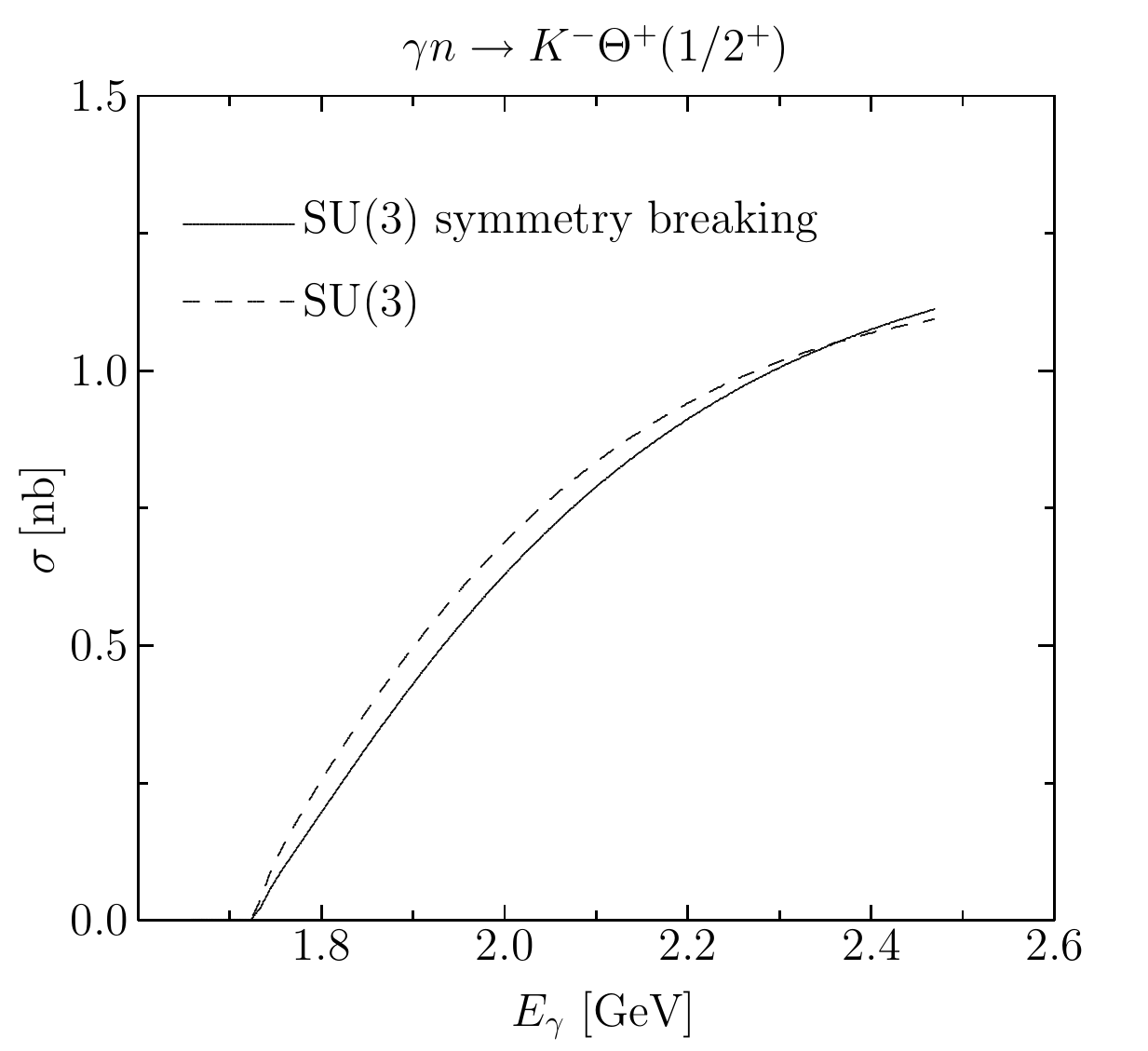}
\includegraphics[width=7.5cm]{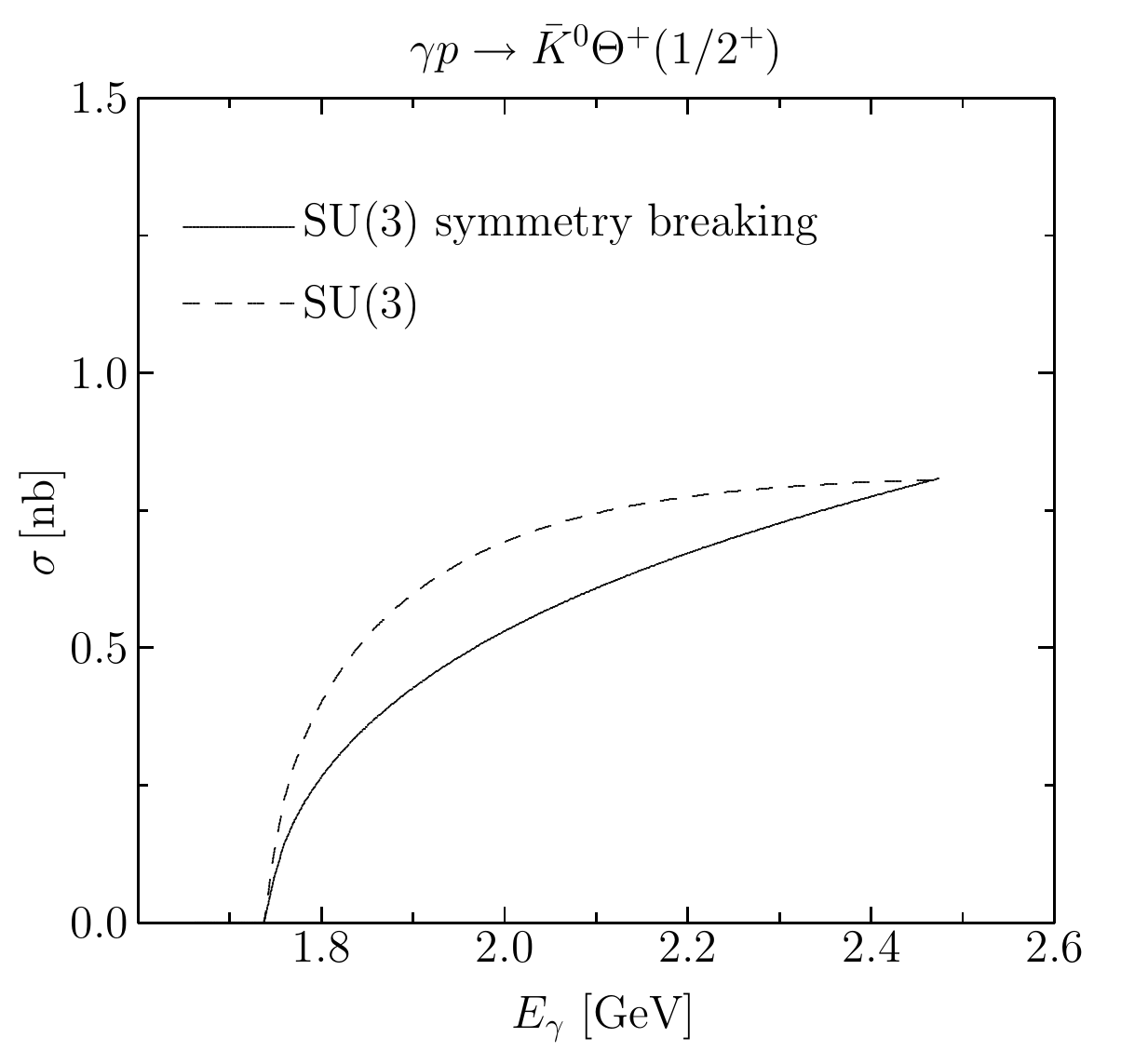}
\end{tabular}
\begin{tabular}{cc}
\includegraphics[width=7.5cm]{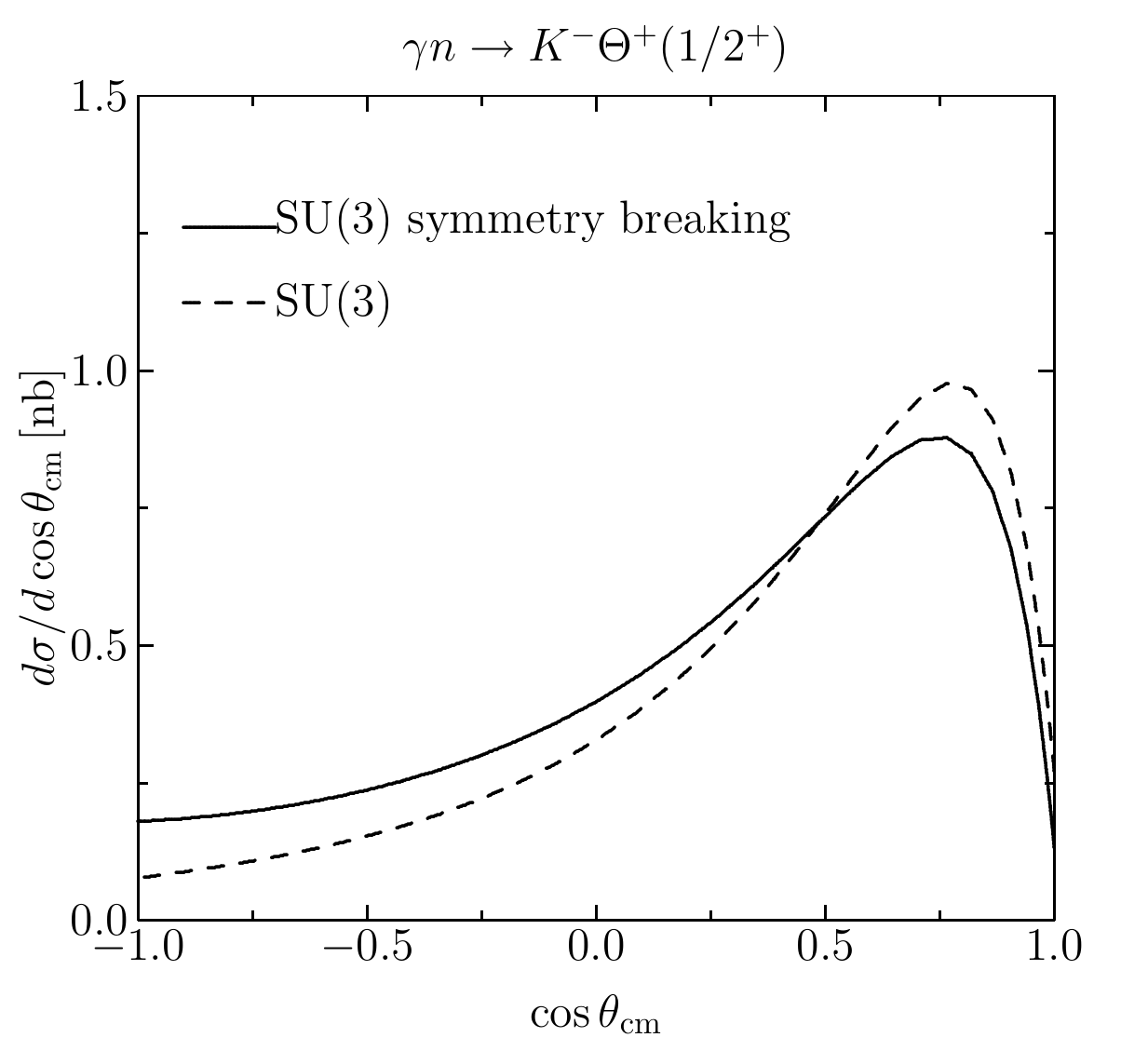}
\includegraphics[width=7.5cm]{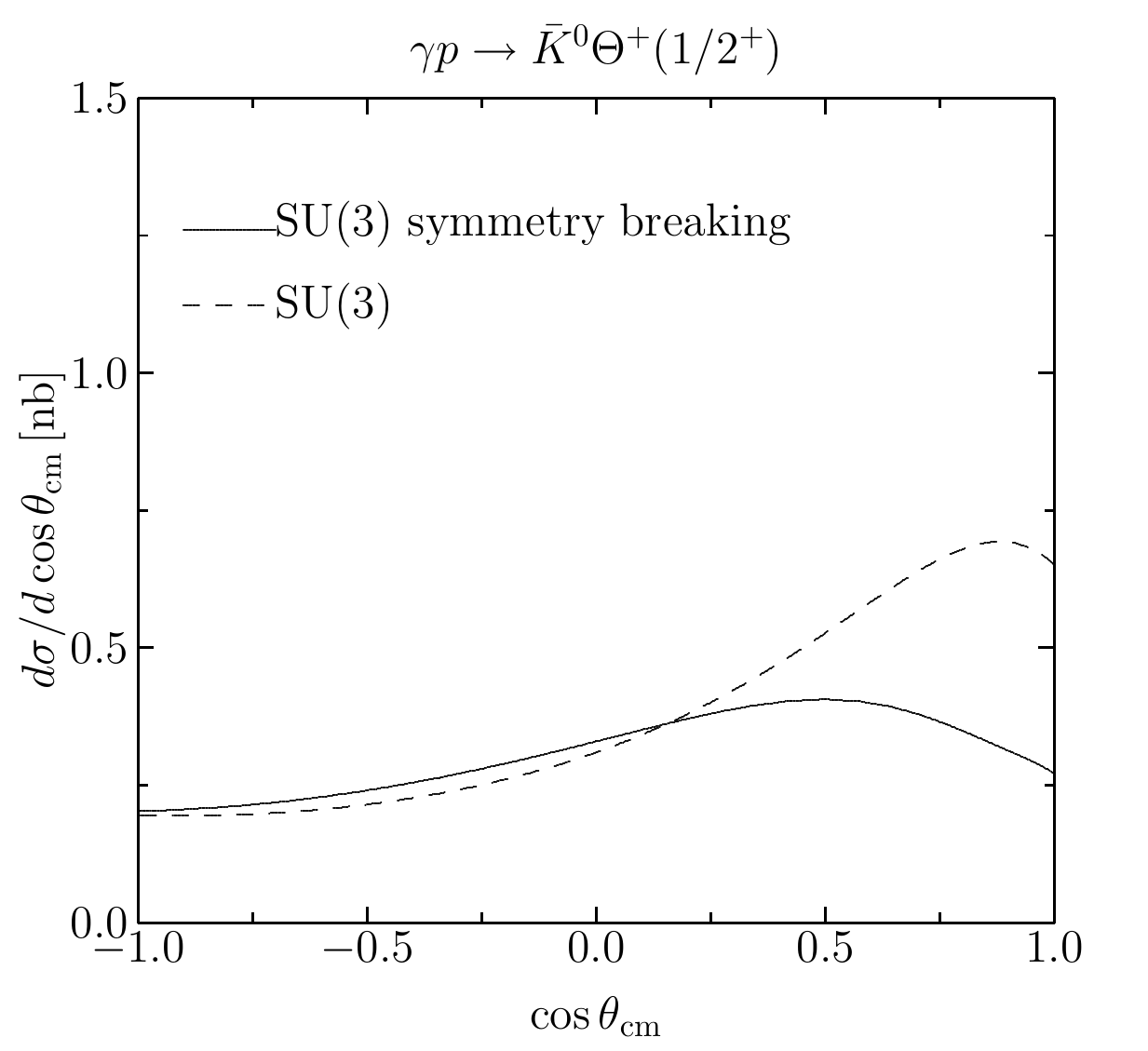}
\end{tabular}
\begin{tabular}{cc}
\includegraphics[width=7.5cm]{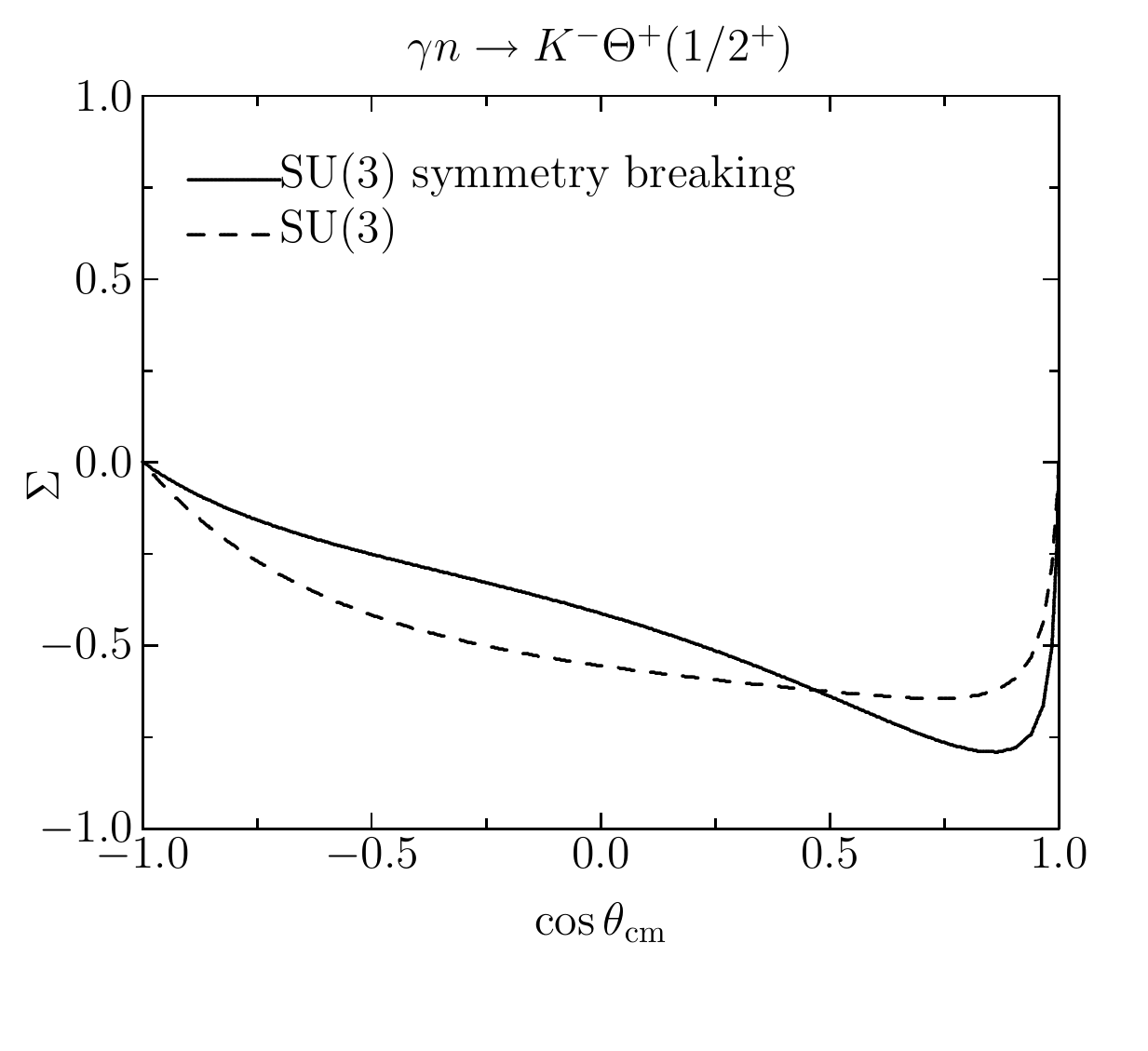}
\includegraphics[width=7.5cm]{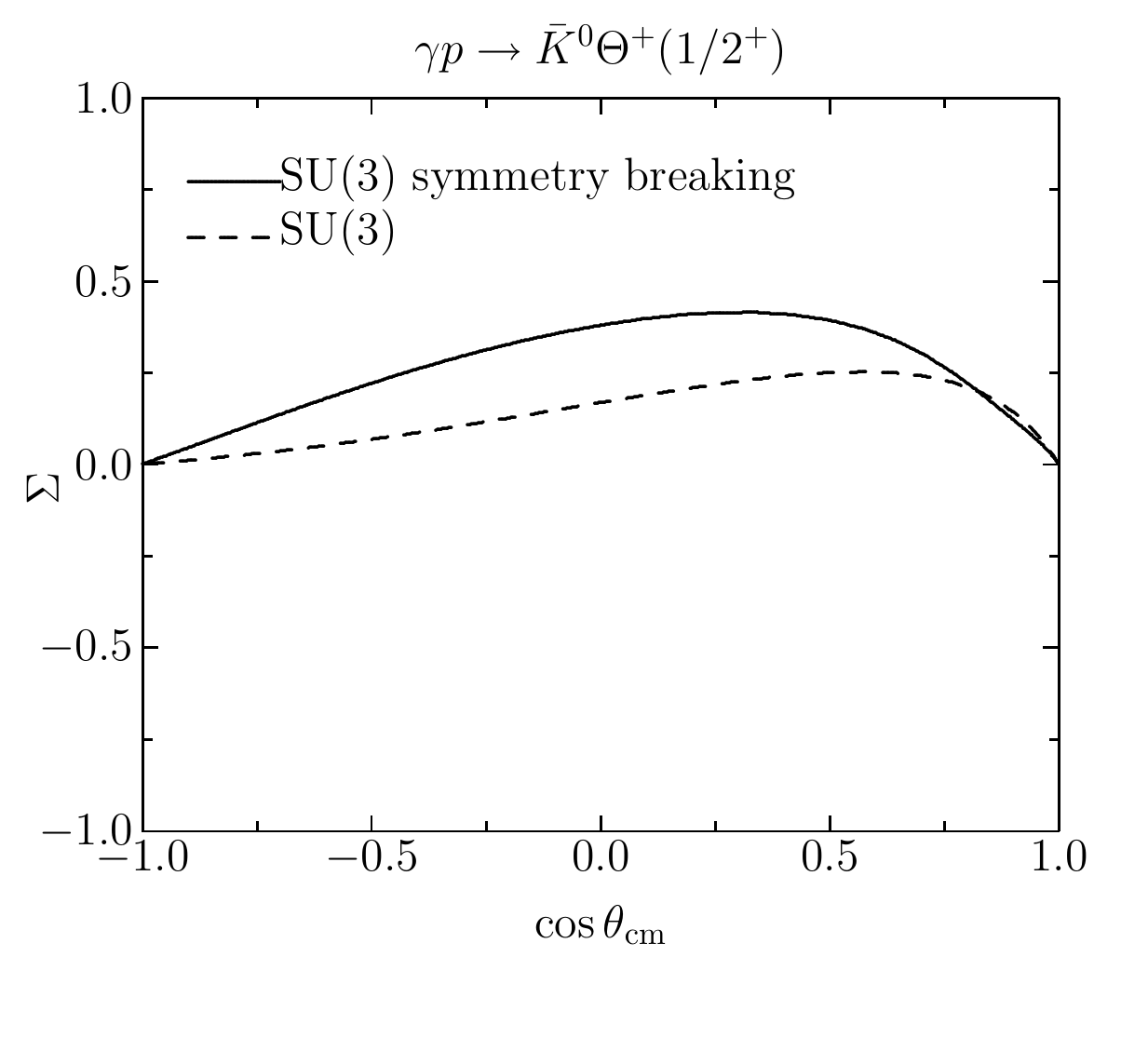}
\end{tabular}
\caption{Effects of SU(3) symmetry breaking on the total ($\sigma$,
upper), differential  ($d\sigma/d\cos\theta_\mathrm{cm}$, middle)
cross sections, and photon-beam asymmetries ($\Sigma$, lower) as
functions of the photon energy $E_\gamma$. The left panels represent
those for the $\gamma n \to K^- \Theta^+$  reaction, while the right
panels those for the $\gamma p \to \bar{K}^0 \Theta^+$. The solid
curve draws the total cross section with SU(3) symmetry breaking,
whereas the dashed one depicts that without it. The curves are drawn
for three different photon energies $E_\gamma$, $2.1$ GeV, $2.2$
GeV, and $2.3$ GeV.} 
\label{fig:4}
\end{figure}

In Sec.~II, it was mentioned that the vector coupling constant 
$g_{K^*N\Theta}$ vanishes in the SU(3) symmetric case due to the
generalized Ademollo-Gatto theorem.  Moreover, the tensor coupling
constant $f_{K^*N\Theta}$ is also very sensitive to SU(3) symmetry
breaking as shown in Table~\ref{tab:coup}.  Thus, it is of great
interest to see the effects of SU(3) symmetry breaking on the
$\Theta^+$ photoproduction. In Fig.~\ref{fig:4}, we show the total
(upper), differential (middle) cross sections, and photon beam
asymmetries (lower) for the neutron and proton targets in the left and
right panels, respectively.  Although the values of the vector and
tensor couplings for the $K^*N\Theta$ vertex are rather sensitive
to the effects of SU(3) symmetry breking, all the results indicate
that SU(3) symmetry breaking does not play any significant role in
describing the $\Theta^+$ photoproduction. The reason can be found in
the fact that while the tensor coupling constant $f_{K^*N\Theta}$ is
almost three times reduced by SU(3) symmetry breaking, the vector
coupling constant $g_{K^*N\Theta}$ comes solely from the wavefunction 
corrections that are also a part of the SU(3) symmetry breaking
effects. Thus, the finite value of the $g_{K^*N\Theta}$ makes up for
the reduction of the $f_{K^*N\Theta}$, so that the effects of SU(3)
stmmetry breaking turn out to be rather small.    
 
\section{Summary and Conclusion}
In the present work, we have investigated the $\Theta^+$
photoproduction, taking the new results of the chiral quark-soliton
model~\cite{Ledwig:1900ri,Ledwig:2008rw} into account.  We first
have briefly reviewed the formalism of the chiral quark-soliton model
for deriving the coupling constants and form factors for the
$K^*N\Theta$ and $KN\Theta$ vertices.  We made use of these coupling
constants and form factors as numerical inputs for calculating the
$\gamma n \to K^- \Theta^+$ and $\gamma p \to \bar{K}^0 \Theta^+$
scattering processes.  

We have examined the effects of each contribution to the total and
differential cross sections.  It turned out that the $K$-exchange
contribution is the most dominant one except for the near-threshold
region in which the $K^*$-exchange contributes mainly.  The
differential cross section has a large bump structure in the forward
direction, which is obviously due to the $K$- and $K^*$-exchange
contributions.     

Since it is of great importance to understand how the $K^*$-exchange
contribution plays a role in the $\Theta^+$ photoproduction, we
thoroughly have studied the effects of the $K^*$-exchange contribution
in various observables for the $\gamma n \to K^- \Theta^+$ and $\gamma
p \to \bar{K}^0 \Theta^+$ reactions.  It turned out that the
$K^*$-exchange contribution is in general the most dominant one in the
$\gamma p \to \bar{K}^0 \Theta^+$ reaction, since there is no
$K$-exchange contribution for the proton target.  The $K^*$-exchange
contribution to the $\gamma n \to K^- \Theta^+$ reaction is in general
about $30\,\%$ but to the $\gamma p \to \bar{K}^0 \Theta^+$ reaction
it is almost everything, since the $K$-exchange contribution is absent
in this case.  In order to see the effects of the $K^*$-exchange
contribution, we also calculated the photon beam asymmetries. It was
shown that with the $K^*$-exchange contribution the photon beam
asymmetries are very different from those without the $K^*$-exchange
contribution.  In particular, the photon beam asymmetry for the proton 
target is changed from the negative sign to the positive with
$K^*$-exchange turned on.

The effects of SU(3) symmetry breaking are remarkable on the coupling
constants for the $K^*N\Theta$ vertex. The vector coupling constant
$g_{K^*N\Theta}$ vanishes without SU(3) symmetry breaking, {\it i.e.}
$g_{K^*N\Theta}=0$.  The tensor coupling constant becomes
$f_{K^*N\Theta}=2.91$ that is by no means small.  When we switched on 
SU(3) symmetry breaking, the vector-coupling constant became finite
because of the wavefunction corrections.  Moreover, the
tensor coupling constant is reduced about by a factor of three.  Thus,
one may see this large changes in the coupling constants in the
observables.  However, it was found that the effects of SU(3) symmetry
breaking were rather small in all observables we calculated in the
present work.  It indicates that the finite value of the vector
coupling constant makes up for the reduction of the tensor coupling
constant.  Thus, the effects of SU(3) symmetry breaking altogether
turn out to be small.   

Recent KEK and LEPS experiments~\cite{Miwa:2007xk,Nakano:2008ee} have
drawn a conclusion that the $K^*N\Theta$ coupling constants must
be small, since the total cross sections turned out to be tiny for the
proton target, being rather different from the present 
results.  In contrast, the present results for the neutron target is
compatible with the experimental data~\cite{Nakano:2008ee}.  In
order to pin down this inconsistency between the theory and
experiments, it is of great importance to have more high-statistics
data for various physical observable for the $\Theta^+$-production
experiment~\cite{private}.  

\section*{Acknowledgment}
The authors would like to thank  J.~K.~Ahn, A.~Hosaka, and T.~Nakano  
for fruitful discussions.  They especially are grateful to T.~Ledwig
for discussing the results from the chiral quark-soliton model.  The
work of S.i.N. is partially supported by the Grant for Scientific
Research (Priority Area No.17070002 and No.20028005) from the Ministry
of Education, Culture, Science and Technology (MEXT) of Japan.  The
work of H.Ch.K. is supported by the Korea Research Foundation Grant
funded by the Korean Government(MOEHRD) (KRF-2006-312-C00507). This
work was done under the Yukawa International Program for Quark-Hadron
Sciences. The numerical calculations were carried out on MIHO at RCNP
in Osaka University and YISUN at YITP in Kyoto University.  
 
\newpage
\end{document}